# Repo Haircuts and Economic Capital:
# A Theory of Repo Pricing

**Wujiang Lou**[1]

**1st draft February 2016; Updated June, 2020**

**Abstract**

A repurchase agreement lets investors borrow cash to buy securities. Financier only lends to securities' market value after a haircut and charges interest. Repo pricing is characterized with its puzzling dual pricing measures: repo haircut and repo spread. This article develops a repo haircut model by designing haircuts to achieve high credit criteria, and identifies economic capital for repo's default risk as the main driver of repo pricing. A simple repo spread formula is obtained that relates spread to haircuts negative linearly. An investor wishing to minimize all-in funding cost can settle at an optimal combination of haircut and repo rate. The model empirically reproduces repo haircut hikes concerning asset backed securities during the financial crisis. It explains tri-party and bilateral repo haircut differences, quantifies shortening tenor's risk reduction effect, and sets a limit on excess liquidity intermediating dealers can extract between money market funds and hedge funds.

**Keywords**: repo haircut model, repo pricing, repo spread, repo formula, repo pricing puzzle.

A repurchase agreement (repo) is an everyday securities financing tool that lets investors borrow cash to fund the purchase or carry of securities by using the securities as collateral. In its typical transaction form, the borrower of cash or seller sells a security to the lender at an initial *purchase price* and agrees to purchase it back at a predetermined *repurchase price* on a future date. On the repo maturity date $T$, the lender (or the buyer) sells the security back to the borrower. The security's settlement agent will cross the security and cash on the same day, in an operational mode of delivery vs payment (DVP) to close the repo trade.

[1] The views and opinions expressed herein are the views and opinions of the author, and do not reflect those of his employer and any of its affiliates.

From the lender's perspective, repo is a secured loan. The initial purchase price is its loan principal, and the repurchase price is principal paid back plus interest earned on the loan. The purchase price is often different from the security's then market value, reflecting a price cushion the lender has demanded to facilitate risk management. The cushion is expressed in two different but interchangeable ways. As a discount to the market price, it is a haircut, denoted as $h$, so that the initial purchase price is the product of 1 minus haircut and the market price. Denote the market price of the security per unit as $P$, the loan amount is $(1-h)P$ with one unit of security as collateral. The amount of $hP$ is a price cushion or price margin in its formal term. The other way around is the margin (advance) ratio denoted by $\eta$, defined as the ratio of the market price of the collateral to the loan amount minus 1. Obviously, $\eta=h/(1-h)$, and, $(1-h)*(1+\eta)=1$. When $h$ is small, e.g., 5%, $\eta$ is very close to $h$, $\eta \cong h$.

As is a loan, the repurchase agreement bears an interest rate, either explicitly stated in the transaction document or implied from the difference of final repurchase price and the initial purchase price. The latter normally associates with one period repo when rate is a fixed number, and the former involves multiple payment and recalculation periods that necessitate floating rate resets. We use $r_p$ to denote the repo interest rate.

A more accurate characterization of repos is secured *margin* loans. Margin or margining is a standard mechanism that maintains the initial cushion against market price fluctuations. Typically this is a daily process that asks the borrower to post additional collateral when the security experiences a price decline. Additional collateral could be cash, more shares of the same security, or other permitted securities. Should the market price increases, reverse margin happens when the lender returns previously held margins, posts cash, or returns a portion of the purchased securities in an operation dubbed as free delivery. In the repo market, daily margin is the standard[2].

As a margined and secured debt instrument, the borrower becomes the repo issuer. A corporate debt is priced by debt interest rate or yield. Repo pricing is atypical of conventional debt instruments, in that haircut also needs to be determined. In fact, repo transaction negotiation covers simultaneously both the haircut and the repo rate. After initial terms are set, such as the principal amount or units of the security and the maturity, a trader at a dealer-bank needs to respond with a haircut and a repo rate (or spread), when he is approached by end-users or investors for their financing requests. Traders and investors alike will need pricing models to procced, especially considering the wide spectrum of different asset classes and repo terms, and the lack of a broadly available, existing market pricing mechanism[3].

[2] In the Treasury GC (general collateral) market, intra-day margin could happen, which correlates to the fact that there is no haircut, i.e., no initial margin to start with.

This article takes up the question of how a repo transaction is priced, including its haircut and interest rate. While repo studies have been an intense academic interest, published works are primarily concerned with understanding repo's role in the short term wholesale funding market and its financial stability implications. Some market surveys are starting to touch upon the topic of repo pricing. At transaction level, there is no known published pricing model that answers this question.

Treating a repo transaction as a credit product with embedded derivatives in the underlying security, we set out to build a joint model of counterparty credit spread and collateral security price dynamics. We combine conventional credit risk management approach with well-industrialized derivatives pricing approach: the former allows us to establish a target haircut level based on credit risk measurements such as probability of default (PD), expected loss (EL), or unexpected loss (UL, also referred to as economic capital, EC), and the latter determines the fair repo rate given the haircut. This combination leaves room for haircuts to be negotiated with end-users while getting compensated through the fair repo rate. From the investors' end, since they will have to finance the residual or haircut portion of the security market value, the overall funding cost (all-in rate) offers the investor an optimal opportunity, so that the dual pricing measures of haircut and repo rate are uniquely determined.

We find that given the existence of haircuts, repo's principal loss is expected to be very small, such that conventional credit risk pricing models are not able to produce any meaningful repo spreads. The very existence of haircuts and the failure of standard credit risk models to explain repo spreads can be termed a *repo pricing puzzle*. Recognizing the systemic nature of unhedgeable and undiversifiable repo losses, we reason that an economic reserve is needed. We introduce a shadow capital account in the Black-Scholes-Merton option pricing economy that associates to the value-at-risk of the hedging errors. This shadow account is financed by some agents who warehouse the risk and charge a shadow cost of capital, which then has to be recouped through the repo return, driving up repo spread. This capital charge component far dominates the traditional expected loss measure, thus solving the pricing puzzle.

[3] Such a market pricing mechanism only exists in the general collateral (GC) Treasury repo market, where haircuts are zero and market participants bid and ask funding rates in a similar manner to the stock market.

Another contribution of this article is identifying economic capital as the mechanism through which haircuts and repo pricing are linked. In particular, these dual pricing measures of the repo pricing can be reduced to economic capital alone, in that the haircuts can be designed to satisfy certain prescribed EC criteria, while the repo spread is driven by the residual economic capital after the haircut has been applied. Measured by the expected shortfall (or tail loss), EC is found to be negatively linear in haircuts. The fair repo spread formula is then approximately linear in haircut as well.

With a transaction pricing model at hand, we also attempt to explain a number of bilateral and trilateral repo stylized facts reported in the literature. The model can quantify tri-party and bilateral repo haircuts differences, show tri-party haircuts' insensitivity to counterparties, set a limit on the funding liquidity generated by dealers' intermediating between collateral rich hedge funds and cash rich money market funds, and corroborate shortening repo maturity as an effective way of lending risk aversion in the wholesales funding market.

The rest of the paper is organized as follows. Section 1 conducts a brief literature review focusing on those more relevant to repo transaction pricing or valuation. Section 2 introduces the repo haircut model with its main components in credit risk measurement targets, collateral asset price dynamics, and solution procedures. Section 3 utilizes the haircut model to explain the difference between bilateral repo haircuts and tri-party repo haircuts, and conducts an empirical study by replaying the haircut hike on asset backed securities during the 2008 financing crisis. Section 4 presents the economic capital's role in the break-even repo rate formula, and synthesizes repo pricing from its dual measures of haircuts and repo rates. Section 5 provides further discussion on economic capital. Section 6 concludes.

## 1. Literature Review

Repo's role in leading to the demise of several major financial institutions and near collapse of the financial system in 2008 has attracted academic research, regulatory, and industry interests. Gorton and Metrick (2012) present evidence that repo haircuts increased dramatically in the US bilateral repo market starting late 2007, especially those concerning securitization products. A run on repo ensured and contributed the crisis. The repo run, however, is not found in the similarly sized tri-party repo market where the repo haircuts barely moved and repo financing for private label securitization was of very limited size (Krishnamurthy, Nagel, and Orlov 2014). Copeland, Martin and Walker (2014) confirm that money market funds (MMF) as the main cash lenders in the tri-party market tend to shut down lending completely rather than asking for higher haircuts in times of stress. Lacking sophisticated analytical tools to determine haircuts, MMFs usually sign up dealer offered haircut schedules. The run in the bilateral market, MMFs' shut down in the tri-party market, and closure of other short term wholesale funding channels such as asset-backed commercial paper (ABCP) conduits, hit hard the few most vulnerable dealers including Bear Stearns and Lehman Brothers, whose subsequent collapses caused systemic distress.

It has since become a contemporary research topic to explain exactly how or why funding market instability such as a repo run could happen. Brunnermerer and Pederson (2009) takes haircut as a speculator's required trading capital, and link market illiquidity defined as the difference between a security's market price and its fundamental value directly to the shadow cost of margin capital. A destabilizing "margin spiral" could develop as the security financier sets haircut based on his knowledge of the fundamental value, the price volatility, and market liquidity.

Aiming specifically at modeling the repo run, dealers' role as funding intermediaries between cash rich MMFs and collateral rich hedge funds (HF) has been studied in the market equilibrium setting. Martin, Skeie and von Thadden (2014) build a dynamic equilibrium model that exploits tri-party and bilateral repo market microstructures (e.g. tri-party daily unwind) to explain the difference between tri-party and bilateral repo haircuts and explores market conditions leading to repo market instability. Infante (2019) focuses on the effect of a dealer default while facilitating collateral movement and extracting desirable excess funding liquidity in a three agent economy (from HF to dealer to MMF) through two chained repos (repo rehypothecation). Assuming that the dealer has zero recovery and the MMF gets to set haircut terms at the extreme downside risk, a market equilibrium then shows that high risk dealers could succumb more easily to a run of collateral from HFs than a run of cash from MMFs.

Obviously market equilibrium models do not aim at transaction pricing: coming up with a repo haircut and repo spread, given a repo trade with a set of terms. Econometric studies could help, especially those conducted at the transaction level. Krishnamurthy *et al* (2014) parse MMF and securities lenders' SEC quarterly regulatory fillings during the period of January 2007 to June 2010. Copeland *et al* (2014) have access to a Federal Reserve Bank's daily collected, confidential tri-party dataset that covers all major tri-party players including dealers and banks, thus a larger dataset than Krishnamurthy *et al*. Their data, however, are aggregates at dealer/investor and collateral type levels, without transaction level details. Hu, Pan and Wang (2019) extract a similar, but more extensive Tri-party dataset from MMF reports, including trade level repo terms, which enables the authors to explore collateral concentration's role in the tri-party repo pricing.

Tri-party repo data and statistics, whilst more readily available, may not bear much relevance to the more dynamic yet opaque bilateral repo market where less liquid, lower credit quality bonds are more likely to be accepted. Krishnamurthy *et al* (2014), for example, report that during the financial crisis, MMF turned away private label ABS papers, but Copeland *et al* (2014) show that such papers are still accepted as tri-party collateral, probably by non-MMF investors such as banks. Indeed, these private ABS are the focus of a confidential dealer-bank's bilateral repo dataset Gorton and Metrick (2012) have gained access to. Bilateral repo trades done by banks and dealers at the time were not subject to regulatory reporting requirements and as a result there is a severe lack of available data. Another limited and yet also inaccessible bilateral repo dataset exists with regard to MBS/ABS, see Auh and Landoni (2016). The data came from a single, large hedge fund that had under management multi-billion securitization assets, financed with repos done with many dealer-banks on the street. The dataset contains some trades that funded different tranches of the same securitization at about the same time, which allows a direct assessment of credit quality (lower tranches are of poorer credit quality) for the first time.

Recognizing the data deficiency in bilateral repos, the Fed launched a bilateral repo data collection pilot project in 2015. The dataset includes broker-dealer entities of 9 major US bank holding companies (BHC), but lasts only a single quarter, is not made public, and may not be representative of the bilateral repo market[4]. Breach and King (2018) collect securities financing data from the Fed's Senior Credit Officer Opinion Survey, for an eight year post-crisis period on a broader range of collateral types. They are able to isolate and focus on risky collateral financing between dealers and their clients (rather than interdealers). Their dataset doesn't seem to contain transaction level details.

[4] Baklanova, Caglio, Cipriani and Copeland (2019) estimate the dataset sheer size accounting for about half of the all bilateral repos. Typical BHC organizes their banks and securities entities separately, which may have their own repo desks. The bank side's repo desk is usually entrusted with deploying the bank's cash as investment through repo, while the securities firm's repo desk handles day-to-day liquidity and securities lending (borrowing securities other desks want to short.) Baklanova et al show the dominance of treasuries and other liquid government or agency bond collateral, a sure sign that the dataset comes from liquidity management desks. For example, there are private label mortgage backed securities (MBS) or ABS collateral. The other half could be more diverse and valuable for the purposes of studying bilateral repo pricing.

In terms of repo pricing, researchers generally agree that collateral quality, volatility, counterparty, and market liquidity affect haircuts and repo rates. Tri-party repo haircuts are known to be lower than bilateral haircuts on the same collateral asset types, some referred to as the haircut difference puzzle (Copeland *et al* 2014). Hu *et al* (2019) finds that neither the tri-party repo haircuts nor repo spreads are sensitive to dealer borrowers. In the bilateral segment, Auh and Landoni (2016) show that both repo haircuts and repo spreads increase as collateral quality deteriorates. Breach and King (2018)'s data also show bilateral repo rates or repo spreads tend to be stable or move together.

The financial stability of the short term wholesale lending market is a priority to regulatory bodies. Bank for International Settlements (BIS) Committee on the Global Financial System conducted a market study (CGFS 2010) on how market participants set credit terms for bilateral repo style transactions. They find diverse market practice in tightening or relaxing securities financing terms, including varying haircut levels, shortening repo tenors, altering counterparty credit limits, restricting collateral asset eligibility, and rejecting certain counterparties. The Financial Stability Board (FSB) enlisted strengthening oversight and regulation of shadow banking as a major task and published a final document on the regulatory framework (FSB 2015). The new framework establishes qualitative and minimal standards for collateral haircuts and governance structures. Although some of standards could be useful in guiding the design and development of transaction repo haircut models, they are not model *per se*.

In the industry, securities financing businesses have been adapting to measures of reforming the financial system, including supplemental leverage ratio, liquidity coverage ratio (LCR), and net stable funding ratio (NSFR). In bilateral repos and bilaterally negotiated, tri-party settled repos for non-government collateral, repo tenors are on average longer than what used to be pre-crisis; most extend beyond 3 months, often with evergreen features[5]. Repos with one year tenor or longer are emerging products for commercial and investment banks and insurance companies -- net cash investors which treat them as a form of short to median term investments.

[5] As an example, a repo is called '4/3/4' evergreen, meaning that the original repo term is 4 months and that with 3 months remaining, it can be extended, i.e., closed out and a new 4 month term repo is entered. If one party does not agree to the extension, it will run off the remaining 3 months. Popular evergreens include '6/3/6', '9/6/9' and '12/9/12'. BASEL LCR requires coverage of a 30 calendar day liquidity stress scenario and 1 year time horizon of NSFR.

Customized transactions are increasingly popular in what are dubbed as structured repos. In collateral upgrade trades (or collateral swaps), for example, the parties' haircut differentials drive the economics of the trades. Dynamic haircuts designed to delever the trades are still rare but not impossible. Meanwhile, broker/dealers and banks are required to fair value repos placed in the trading book, with repo counterparty credit risk explicitly measured and managed (BCBS, 2016), in a way not dissimilar from OTC derivatives. Lengthened tenors, new structured features, and fair value requirement all necessitate consideration of counterparty credit risk and interaction between haircuts and borrower credit, the main subject of this research.

To accommodate these developments, a robust modeling capacity of repo transactions becomes a pressing need, especially given "the absence of a clear understanding of the constitution of haircuts/initial margins" (Comotto, 2012). Different from the literature surveyed above, our primary motivation is to provide a pricing model for repo traders and investors as well to price a repo transaction. The general equilibrium approach does not offer transaction level pricing. Empirical studies could in theory come up with regression models that allow a transaction to be priced. Such regression models are obviously limited by the private nature of bilateral repo transactions, which have made it impossible for large scale data collection and data sharing[6].

Second, our focus is repos with non-government securities as collateral. Treasuries repos have well established market and trading mechanisms that have not only afforded in depth research efforts but also allowed traders to look up to the market to quote a repo. Repos involving non-government securities collateral are much less understood, mainly because of their over-the-counter (OTC) nature. US treasuries' liquidity is also superior to other collateral. In fact, it is fair to say that treasuries repos are not so much as debt investment instruments as liquidity products. Our proposed model does apply to treasuries repos, but its impact is limited, because of the reasons cited above.

## 2. Repo Haircut Model

In the financial market, haircut is a discount on the market value of a financial asset when used as collateral for a financial obligation or debt[7]. In the early days, stock loan brokers used heavy price discount to withstand stock market meltdowns. Intuitively, a haircut on a stock can be taken as the worst daily or weekly price decline, depending on how long the financer thinks it will take to liquidate the stock to recover his loan. According to an industry survey (Comotto 2012), rules of thump based on market experience and simple price volatility measures, such as a multiple of price return standard deviations, are popular ways of computing haircuts. As risk management advances, value-at-risk (VaR) measure has been adopted to arrive at a VaR-based haircut, e.g., the 99% tail loss during a 10 day period over a sufficiently long observation period.

[6] The data sources cited here are all confidential or proprietary.

[7] Ashcraft et al (2010) find a quote that shows taking discount on collateral value long existed 2000 years ago.

VaR-based haircuts are data-driven. Such an approach is as good as data is, and carries the usual caveat that history may or may not repeat itself. There are cases when pricing history is absent, too short, or long enough but yet to have seen a meaningful market stress. Lou (2017) develops a parametric haircut model by employing a double-exponential jump-diffusion model for collateral market risk and solving haircuts to target credit risk measurements or credit ratings. Computational results for main equities, securitization, and corporate bonds show potential for uses in collateral agreements or situations where counterparty independent haircuts are required, e.g. for regulatory capital calculations, and collateralization for OTC derivatives.

Strictly speaking, these collateral haircuts are not repo haircuts, because their specification has not made use of the borrower's credit. Repo has full recourse to the borrower. A security with a market price of \$100, for example, can be used to collateralize \$85 cash loan, subject to 15% discount or haircut. If the borrower fails to pay back this \$85 loan and the market price has dropped to 80, the lender is \$5 short after selling the security at the market. That \$5 becomes a claim on the lender's estate in the senior unsecured rank.

Like holders of other debt instruments, a repo lender is therefore exposed to the borrower's default risk. As a secured debt, repo has a limited exposure amount, far less than the full loan amount. For overnight repos with zero haircut, the exposure is one day price decline. For term repos (i.e., non-overnight repos), the daily margining mechanism kicks in which keeps the exposure essentially one (future) day, regardless of the repo tenor. By design, repo has only one day market risk exposure contingent on borrower default.

Complications do arise from practicalities evidenced by margining procedures and default settlement provisions embodied in MRA (Master Repo Agreement) or GMRA (the Global Master Repo Agreement). A grace period of at least a day is given, for example, to allow a counterparty to amend a failed margin call. A party can raise a pricing and margin calculation dispute that the provisions provide a time window to resolve. In all likelihood, the period from the time when a party last meets a margin call to the time when collateral asset sales are completed will be more than one day.

Formally known as a margin period of risk (MPR), this period could cover a number of events or processes, including collateral valuation, margin calculation, margin call, valuation dispute and resolution, default notification and default grace period, and finally time to sell collateral to recover the lent principal and accrued interest. Obviously, a longer MPR directly increases repo's exposure. If MPR is zero, repo has no exposure, no risk.

As critical as MPR is to repo's risk, it is not a number readily available or can be rigidly derived. Typical MPR ranges from 5 days to 20 days, depending on collateral asset liquidity, concentration, and contractual factors (Andersen, Pykhtin, and Sokol, 2017). In regulatory risk capital related works, a 10-day MPR is taken as the standard reference for bilateral OTC derivatives and security financing transactions. Particularly for tri-party repos, because the third party custodian standardizes and takes over collateral valuations and certain aspects of default settlement, the length of MPR is expected to be shorter than that of bilateral repo MPRs.

In our model, MPR is taken as a known input. Repo exposure is based on the negative price movement during the MPR, assuming margin is perfected maintained at the beginning of the MPR, and collateral security is sold at the period end market price.

### 2.1. Repo haircuts as a credit risk mitigation measure

The lender's nominal exposure during the MPR is principal plus accrued and unpaid interest up to the default event date. For haircut modeling purposes, we assume no lapse between the last margin date and the default event date, so that the repo exposure in the MPR is flat. Any shortfall from the sales proceeds to cover the exposure results in an unsecured claim which is *pari passu* to the borrower's senior unsecured obligations and subject to the same recovery process.

Repo-style securities financing transactions are kin to OTC derivatives: they have the same exemption from the automatic stay in the bankruptcy courts, hold the same unsecured rank, and are subject to the same provisions on bilateral counterparty credit risk in the regulatory risk capital space (BSBC 2016). To be clear, our haircut model considers borrower's default risk only, typical of loan analytics when the lender itself is a going concern. This also aligns with BASEL III's standard on counterparty valuation adjustment (CVA).

Suppose a hypothetic bank B and a client C enter a repo transaction, where B lends $M(t)$ amount of cash to client C on $\Delta(t)$ units of collateral security with a price process $B(t)$. In typical industry jargon, B enters into a *reverse* repo, while C has a repo.

At a constant haircut $h$, repo margining ensures $M(t)=(1-h)\Delta(t)B(t)$, for $t<min(T, \tau)$, $\tau$ is the default time of the counterparty and $T$ is repo maturity. Let $u$ be the MPR, a known and fixed time period, then $\tau+u$ is when the security is completely sold. For simplicity, we assume all shares are sold at time $\tau+u$ when the mid market price is $B(\tau+u)$.

The actual selling price contains a discount $g$, $1>g\geq 0$, to the mid, which takes into account the bid/ask spread or other market liquidity considerations, such as a block size discount or a fire sale discount. Brunnermerer and Pedersen (2009) define market liquidity as the difference between security transaction price and its fundamental value, an endogenously generated measure. Here we treat $g$ as an external input, assessed by traders based on collateral asset market condition. Market liquidity is commonly cited as a contributing factor to haircut, e.g., Gorton and Metric (2012), Krishnamurth et al (2014), Copeland et al (2014), Martin et al (2014).

Incorporating the liquidity discount, the sales proceed becomes $\Delta(\tau)B(\tau+u)(1-g)$. Party B's residual exposure to party C in the senior unsecured rank is $(M(\tau)-\Delta(\tau)B(\tau+u)(1-g))^+$. Denote $R_c$ as C's recovery rate, $\Gamma_c(t)$ C's default indicator, 1 if $\tau\leq t$ , 0 otherwise. We write B's loss process at time $t$ as follows,

$$L(t) = L_{gd}\Gamma_c(t)\Delta_\tau\left(B_\tau(1-h) - B_{\tau+u}(1-g)\right)^+, \tag{1.a}$$

Or in a differential form,

$$dL(t) = L_{gd}(1-g)\Delta_t B_t\left(\frac{1-h}{1-g} - \frac{B_{t+u}}{B_t}\right)^+ d\Gamma_c(t) \tag{1.b}$$

where $L_{gd}$ is the loss given default, applied to reflect the repo's recourse on borrower C. For non-recourse repos (rare), one can simply set to 1. Obviously, if $u=0$ and $g=0$, there is no repo loss.

The loss function reflects the credit enhancement role played by haircutting. Let $y = 1 - \frac{B_{t+u}}{B_t}$ be the price decline. If $g=0$, $Pr(y>h)$ equals to $Pr(L(u)>0)$. Trivially, there will be no loss, if the price decline is less or equal to $h$. A first dollar loss will occur only if $y>h$. $h$ thus provides a cushion before a loss is incurred. Given a target rating class's probability of default (PD) $p_0$, the first loss haircut can be written as

$$h_p = \inf \{h \in R^+ : Pr(L(u) > 0) \leq p_0\}. \quad (2)$$

For rating agencies that employ expected loss (EL) based target per rating class, haircuts are based on EL target $L_0$,

$$h_{EL} = \inf \{h \in R^+ : E[L|h] \leq L_0\}, \quad (3)$$

$L_0$ can be set based on criteria of certain designated high credit rating, whether bank internal or external such as Moody's 'Aa1' rating. Apart from PD and EL, another common measure adopted in credit risk management is credit value-at-risk (CVaR). Given a quantile $q$, CVaR is defined as follows,

$$CVaR_L(h) = \inf \{l \in R^+ : \Pr(L(T) > l|h) \leq 1 - q\}, \quad (4)$$

A typical value of $q$ is 99.9% for one year holding period, $T=1$.

The difference between CVaR and EL is unexpected loss (UL), a reserve capital measure formally termed economic capital (EC), $EC(h)=CVaR(h)-EL(h)$. The VaR measure is replaced by expected shortfall (ES) in the newly proposed BASEL market risk capital rules (BCBS, 2016)[8], $EC(h)=ES(h)-EL(h)$. Naturally we can define a haircut to minimize capital requirement $C_0$.

$$h_{EC} = \inf \{h \in R^+ : EC(h) \leq C_0\}, \quad (5)$$

where EC is measured either as CVaR or ES subtracted by EL.

A street firm usually considers and organizes repo business from two different perspectives: one is liquidity line of operation and the other is lending line business (see also footnote 4). Treasury repos, especially those with short tenors, generally fall under the liquidity line. These repos do not earn many basis points above benchmark short term interest rates, such as the Fed fund rates. The lending line is subject to the firm's credit risk management practice and policies. Repo haircuts, treated as a credit risk mitigation tool, require credit department's approval, often per transaction. A firm's credit policy thus must provide a method of deriving a haircut at the transaction level. In doing so, it may choose a haircut policy. They may choose S&P or equivalent ratings that are based on controlling probability of loss, e.g., `AA` rating with a PD, $p_0=0.00005$. Or they can decide on Moody's or similarly EL based ratings, e.g., `AA` rating with an EL target, $L_0=0.0004$. Following the financial crisis, banks have developed their own internal rating systems and have specifications of these PD or EL criteria per rating.

[8] Expected shortfall has yet to make it into the BASEL's credit risk framework, although researches are under way, e.g., Osmundsen, Kjartan Kloster, 2018. "Using expected shortfall for credit risk regulation," Journal of International Financial Markets, Institutions and Money, Elsevier, vol. 57(C), pages 80-93.

To solve haircuts from equations (2, 3, 4 or 5), the probability distribution of $L(T)$ needs to be specified, which is governed by the default time model and the collateral price dynamics provided below.

## 2.2. Credit Spread and Collateral Price Dynamics

The loss function (equation 1.b) has an exposure amount dependent on a forward put option on the collateral asset. Since the strike $\frac{1-h}{1-g}$ is practically less than 1 as haircut is non-zero, the put is out-of-the-money. And because the put expiry -- the MPR $u$ – is very short, the value of the put is predominantly decided by skewness and tail characteristics of the asset return distribution. Asset price models with stochastic volatility and jumps in both return and volatility are shown to produce better empirical results in studies of stock indices (Eraker, Johannes and Polson, 2003). Indeed, stochastic volatility models are welcomed in the options pricing literature and industry practice, but they are shown to be less effective for very short term options (Eraker 2004).

The double exponential jump-diffusion (DEDJ) model (Kou, 2002) we choose here is popular in exotic and path dependent options pricing, which allows asymmetric up and down jumps. A single asset's price process $B(t)$ is written as follows,

$$X_t = \log\left(\frac{B_t}{B_0}\right)$$

$$dX_t = \mu dt + \sigma_a dW_a(t) + \sum_{j=1}^{dN_t} Yj \tag{6}$$

where $\sigma_a$ is the asset volatility, $\mu$ the asset return, $dW_a(t)$ a Brownian motion, $dN(t)$ a Poisson process with intensity $\lambda$, and $Y_j$ a random variable denoting the magnitude of the $j$-th jump. With DEJD, $Yj$, $j=1, 2, ...$, are a sequence of independent and identically distributed double-exponential random variables with the probability density function (pdf) $f_Y(x)$ given by

$$f_Y(x) = p_u \eta e^{-\eta x} I\{x \geq 0\} + q_d \theta e^{\theta x} I\{x < 0\} \quad (7)$$

where $p_u$ and $q_d$ are up jump and down jump probabilities, $p_u+q_d=1$. The up jump size is exponentially distributed at a rate of $\eta>1$. Similarly $\theta>0$ is down jump's rate.

The forward put drives the repo's market exposure which is contingent on the borrower's default. In the credit derivatives world, the reduced form model has become the main default modeling approach. Corporate credit spreads are shown to exhibit log OU (Ornstein-Uhlenbeck) behavior (Duffie 2011) which is positive, mean-reverting, and highly elastic in that it allows large moves in credit spread. In the log OU model, the default intensity $\lambda_c(t)$ for the default time $\tau$ is written as follows,

$$\lambda_c(t) = \mathrm{e}^{y(t)},$$

$$dy(t) = k\big(\bar{y} - y(t)\big)dt + \sigma_c dW_c(t) \quad (8)$$

where $k$ is the mean reversion rate, $\bar{y}$ the mean reversion level, $\sigma_c$ credit spread volatility, and $dW_c(t)$ a Brownian motion defined in a proper probability space $(\Omega, \mathcal{F}, P)$.

The asset price process is correlated with the intensity, $\langle dW_c(t), dW_a(t)\rangle = \rho dt$, $\rho$ the correlation coefficient. $dW_a$ can be written in a factor form, $dW_a = \rho dW_c + \sqrt{1-\rho^2} dW$, where $dW$ is independent of $dW_c(t)$. A negative correlation implies that when the credit deteriorates, the asset return also tends to go down, thus creating a scenario of the "wrong-way risk" (WWR).

The log OU dynamic spread model, while empirically supported, is known for its lack of analytical tractability. A Monte Carlo simulation is commonly needed. Suppose that we simulate a path of $W_c(t)$, $\mathcal{F}_c = \{W_c(t),\ 0 \leq t \leq T\}$. It then leads to a path of $y(t)$, denoted as $y_F(t)$. Conditioning on $\mathcal{F}_c$, $B(t)$ has a changed drift term but otherwise remains a DEJD process, as listed below,

$$\frac{B(T)}{B(t_-)} | \mathrm{F}_c = e^{\alpha(\mathrm{t},\mathrm{T},W_c) + \sigma_a \sqrt{1-\rho^2}(\mathrm{W(T)} - \mathrm{W(t)})} \prod_{i=N(t)}^{N(T)} e^{Y_i}$$

$$X_t(\mathrm{T}) | \mathrm{F}_c = \log\left(\frac{B_T}{B_{t-}}\right) = \alpha(\mathrm{t},\mathrm{T},\mathrm{W}_c) + \sigma_a \sqrt{1-\rho^2}\big(W(T) - W(t)\big) + \sum_{j=Nt}^{N_T} Yj$$

$$\alpha(\mathrm{t},\mathrm{T},\mathrm{W}_c) = \mu(\mathrm{T}-\mathrm{t}) + \sigma_a \rho\big(W_c(\mathrm{T}) - W_c(\mathrm{t})\big). \quad (9)$$

Fix a time horizon *T*, the conditional expected loss is

$$\mathrm{E}[\mathrm{L}(\mathrm{T})|\digamma_c] = L_{gd}(1-g)\int_0^T dP_{y_F}(t)E[\Delta_t B_t|\digamma_c]E[\left(\frac{1-h}{1-g}-\frac{B_{t+u}}{B_t}\right)^+|\digamma_c] \tag{10}$$

where $P_{y_F}$ is the conditional default probability, $P_{y_F}(t) = 1 - e^{-\int_0^t e^{y_F(s)}ds}$.

To compute the tail probability of loss at time *T* exceeding an amount *b*, $P_b=Pr(L(T)\geq b)$, we again resort to conditioning to arrive at,

$$P_b = \mathrm{E}[\mathrm{I}\{\mathrm{L}(\mathrm{T}) \geq b\}] = E[\mathrm{E}[\mathrm{I}\{\mathrm{L}(\mathrm{T}) \geq b\}|\digamma_c]],$$

$$E[\mathrm{I}\{\mathrm{L}(\mathrm{T}) \geq b\}|\digamma_c] = \int_0^T dP_{y_F}(t)\Pr\{X_{t+u} \leq \log\left(\frac{(1-h)L_{gd}-\frac{b}{M_t}}{(1-g)L_{gd}}\right)|\digamma_c\} \tag{11}$$

where $b < L_{gd}(1-h)\Delta_t B_t$. $\frac{b}{M_t}$ is the relative loss measured against the repo principal $M_t = (1-h)\Delta_t B_t$, and *Pr{.}* is the cumulative density function (cdf) of $X_t$.

Repo style security financing transactions operate either with fixed positions where $\Delta(t)$ is constant or constant exposure where $\Delta(t)B(t)$ is constant. The latter corresponds to a constant loan amount $M_0$, which is the norm in repo. The former is typical of a sec lending or total return swap (TRS) funding transaction[9]. Since our focus is repo, we can take out $E[\Delta_t B_t|\digamma_c]$ in equation (10) to normalize the loss as the percentage of the loan. Accordingly, in the remaining text, we assume $M_t=M_0=1$, a unit notional.

### 2.3. Negative linear VaR and ES in haircuts

The conditional expectation $E[\left(\frac{1-h}{1-g}-\frac{B_{t+u}}{B_t}\right)^+]$ in equation (10) is evaluated by inverse Laplace transform (Cai, Kou, and Liu, 2014), so is cdf of $X_t$ for equation (11). $P_b$ is then obtained by running a Monte Carlo (MC) simulation on equation (11). EL is obtained similarly from simulating equation (10). Zero correlation is a special case where equations (10 & 11) can be evaluated without the need of running MC simulation, once the default probability $P_y$ is computed separately.

[9] As far as the margin account is concerned, the latter is equivalent to the use of the same collateral to fund the margin account, which alternatively could be funded with cash or government debts. This has a negative leveraging effect when B's price declines. Some may consider introducing price floors to limit the extent of this leverage for certain high volatility asset classes. In the simulation model, this is relatively straightforward to capture and will be left for future exercises.

Obviously $P_b$ depends on $h$, better written as $P_{b|h}$. Fixing a haircut $h$, equation (11) gives loss distribution $P_b$ as a function of $b$. We set up a loss grid in $b$ with step size $\Delta b$, $b_0=0$, $b_1$, $b_2$, ..., $b_I$, and $b_i = i\Delta b$. For each $b_i$, $P_{b/h}$ becomes a function of $h$. In particular, for $b_0=0$, the no loss probability $P_{0/h}$ can be inverted to solve for $h$ given a target level of $P_0$ for a given PD based rating class.

It is useful for implementations to note that equation (11) is translational in $h$ and $b$, i.e.,

$$P_{b^*|h^*} = P_{b|h},$$

$$b^* = b + (h - h^*)L_{gd} \tag{12}$$

An implementation could start with $h=0$, choose $\Delta b=L_{gd}\Delta h$, compute $P_{b|h}$ on the loss grid $b_i$. Then on a haircut grid, $h^*=i\Delta h$, $i=0, 1,2, ..., I$, use equation (12) directly to look up $P_{b^*|h^*}$.

For a fixed $h$, $VaR_L$ can be solved by finding $b$ such that $P_{b/h}=1-q$, loss not exceeding $b$ with the confidence interval $q$. For a different haircut $h^*$, we then have $P_{b^*|h^*} = 1 - q$, so long as $b^*$ is determined by (12). This states that $b^*$ is the $VaR(h^*)$, if $b$ is $VaR(h)$. Equation (12) gives

$$VaR(h^*) = VaR(h) + L_{gd}(h - h^*) \tag{12.b}$$

With a sufficiently large haircut, $VaR_L(h)$ (equation 4) could be zero, when $P_0 \geq q$. (12.b) can be refined,

$$VaR(h^*) = \begin{matrix} VaR(0) - L_{gd}h^*, if\ VaR(0) > 0\ and\ h^* < \frac{VaR(0)}{L_{gd}} \\ 0, else \end{matrix} \tag{12.c}$$

Therefore, for a meaningful combination of $q$ and $h$, non-zero VaR of the repo loss decreases linearly as haircut increases. This is not surprising at all, since haircut directly cuts down loss and has the effect of truncating the loss distribution.

Similarly with $b$ fixed, $E[L(T)]$ can be computed as a function of $h$ to solve for haircut by definition (equation 3). In particular, there is also a translational formula[10] for the (conditional) tail loss expectation,

$$E[L_h(T)|L_h(T) \geq b] = \frac{E[L_{h^*}(T)]}{P[L_h(T) \geq b]} + b, \tag{13}$$

$$h^* = h + \frac{b}{L_{gd}}.$$

where we have added subscript $h$ to the loss function $L$ to denote its association with haircut $h$ as defined in equation (1). Setting $b$ to $VaR(h)$ (so that $P[L_h(T) \geq b] = 1 - q$) leads to a convenient formula that computes ES given a $q$-tile,

$$ES_{h^*} = \frac{E[L_{h_q}(T)]}{1-q} + VaR(h^*) \tag{14}$$

$$h_q = h^* + \frac{VaR(h^*)}{L_{gd}}$$

Plugging (12.c) into (14) to arrive at,

$$ES_{h^*} = \frac{E[L_{h_q}(T)]}{1-q} + VaR(0) - L_{gd}h^* \tag{14.b}$$

The same negative linear term in haircut seen in (12.c) now appears in the expected shortfall. This negative linear relationship does not exist in probability of no loss and expected loss, the other two popular measures of credit risk. VaR and ES thus have a unique advantage in repo haircut design.

To compute $ES$ given $q$ and an $h*$, first use (12.c) to get $VaR(h^*)$, $h_q$ from (14), $E[L_{h_q}(T)]$, then (14.b) to get $ES$. $ES$ can of course be computed from the discrete loss distribution obtained from equation (11). Since it depends on the tail of the distribution, its accuracy would require a very fine loss grid $b_i$, but equation (14.b) allows us to circumvent that.

Cai, Kou, and Liu (2014) develops an inverse transform algorithm with both discretization and truncation error controls to promote numerical stability. The controls are separately computed for cdf and options. For our purposes, such an inverse transform is run per path and it will be more efficient to apply the same transform setting to obtain cdf of $X_t$. Lou (2017) revises error controls so that the same truncation and discretization parameters apply to the inversions of cdf and put options. DEDJ model estimations conducted on US main equity index, securitized products, and US corporate bond indices over a 10 year span show reasonable model stability and behavior.

[10] See Appendix A for derivation.

The model discussed above in principal works for securities lending transactions. A sec lender has a loss on the other side of price movement, i.e., when price appreciation over the MPR exceeds the extra margin of $hB_t$ posted by the security borrower. Expected loss of the sec lender would then relate to a call option payoff on a constant unit of the collateral security. This would be left for future research.

## 3. Predicting Repo Haircuts

A haircut model's primary application is to predict repo haircuts in accordance with a bank's securities financing business model and risk management capacity. If the bank treats the repo as a secured loan to be carried on its banking book, it then needs to set the haircut that produces the firm's desired lending profile on a wholesale exposure. Take for example, suppose the firm is comfortable lending out to 'A' rated wholesale clients unsecured, it could trade a reverse repo with a 'BB' rated party, provided that the trade has a haircut designed to make the overall credit risk profile matching that of 'A' rated counterparties.

### 3.1. US Treasury haircuts

The US Treasury securities are the single most dominant collateral in the repo market. Table 1 shows 1 day, 5 day, and 10 day VaR and ES market risk measures taken over a 5 year period from 3/7/2006 to 3/7/2011 with the financial crisis in the middle, and another 5 year period ending in May 2020 when the coronavirus roiled the market in Q1 2020. One day VaR-based haircuts for active 10 year notes and 30 year bonds are 1.41% and 2.60% respectively. If one adopts a MPR of 5 days, notes haircut is 2.9% and bonds 5.22%[11].

These data driven, VaR-based haircuts can be complimented by a parametric haircut model. Estimated DEJD for the current 10 year notes price time series has $(\mu, \sigma_a, \lambda_u, \lambda_d, \eta_u, \eta_d)$ = (-0.014575,

[11] These numbers are surprisingly consistent to typical ISDA Credit Support Annex (CSA) valuation percentages applied to Treasury notes and bonds, 98% and 95%, which imply 2% and 5% haircuts.

0.071804, 27.551, 22.746, 186.42, 232.44) and produces 0.3507 in skewness and 6.1927 in kurtosis, comparing to sample skewness of 0.1532 and kurtosis of 6.4770.

Table 1. VaR (99 percentile Value-at-Risk) and ES (97.5-percentile Expected Shortfall) during two 5 year periods containing the 2008 financial crisis and current coronavirus market distress.

| | Active 5y Notes | | Active 10y Notes | | Active 30y Bonds | |
|---|---|---|---|---|---|---|
| period/days | VaR | ES | VaR | ES | VaR | ES |
| 2008-1d | 0.89 | 0.91 | 1.41 | 1.45 | 2.60 | 2.68 |
| 2008-5d | 1.64 | 1.70 | 2.90 | 2.92 | 5.22 | 5.45 |
| 2008-10d | 2.06 | 2.11 | 3.69 | 3.67 | 7.79 | 7.42 |
| 2020-1d | 0.49 | 0.51 | 0.93 | 1.08 | 2.16 | 2.64 |
| 2020-5d | 1.06 | 1.08 | 2.14 | 2.26 | 4.97 | 5.47 |
| 2020-10d | 1.59 | 1.52 | 2.72 | 2.88 | 6.96 | 6.37 |

Figure 1 shows predicted collateral haircuts targeting Moody's and S&P investment grade (IG) credit ratings, i.e., $h_{EL}$ and $h_p$. The hypothetic S&P 'A' and above rating targeted haircuts are about 1~2 points higher than corresponding Moody's rating targeted haircuts. Caution should be taken, however, these default rates and loss rates are examples and not directly comparable against each other. A firm may choose one haircut policy, e.g., designing haircuts to match an overall 'AA' Moody's rating or some internal credit rating criteria, with the understanding that different rating criteria does lead to some variations in haircut levels, as shown here. We also show $h_{EC}$ or $h_{ES}$ (as ES is used as EC), which produces haircuts somewhat in the middle of Moody's and S&P's rating haircuts.

Numbers shown in Figure 1 are *collateral haircut*, i.e., counterparty independent haircuts. Among the money market fund families' tri-party repos, Treasury haircuts are in a very tight range around 2% (Tables 9 and 10 in Hu *et al* 2019), with a mean remaining maturity of 6.21 years, and notes on average accounting for 79%. Top ten dealers' mean haircuts range from 1.97% to 2.10%, and the standard deviations range from 0.14% to 0.52%. These ranges overlap closely with the above one and 5 day VaR-based collateral haircut range of 1.4% and 2.9%. This makes it difficult to tell whether MMFs are using collateral haircuts for their Treasury repos or they price repo haircuts in a tight range because the dealers' credit quality are in a tight range. We will have to look at more volatile collateral asset classes.

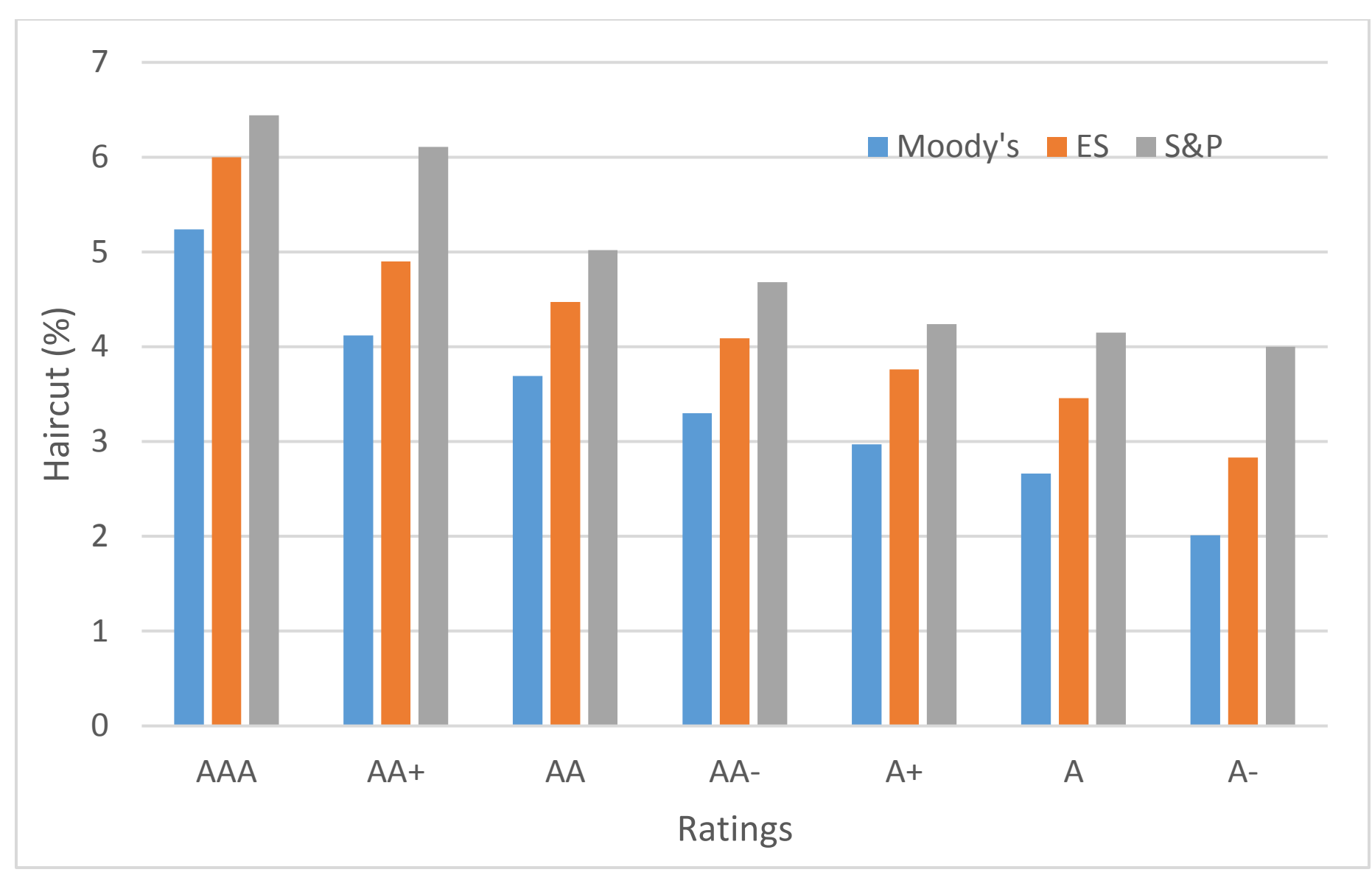

Figure 1. Predicted Treasury notes haircuts (MPR 5 days) targeting hypothetical Moody's one year loss rates per Bielecki (2008), S&P's average one year default rates (Standard & Poor's 2015)[12], and EC-based haircuts.

### 3.2. Equity repo haircuts

Next we consider a bank lends to an 'A' rated wholesale counterparty, for 1 year with US main equities as collateral[13]. The asset model is estimated with a 5 year historical period from 1/2/2008 to 1/2/2013. The borrower's spread dynamics is assumed to have 90 bp initial and mean hazard rate levels, mean reversion speed of 0.5, and spread log OU volatility of 1.50, i.e., $k=0.5$, $\bar{y}=log(0.009)$, $\sigma_c=1.5$, $\lambda_0=0.009$ in equation (6), such that its 5 year CDS prices at 125 bp[14].

The EL based haircut targeting Moody's 'Aa2' without giving consideration to the borrower's credit is 15.53% (of collateral haircut) for an MPR of 10 days. With borrower's default probability considered, 7.98% (Table 2) suffices to reach the same 'Aa2' profile under zero credit and asset correlation. If the correlation is stressed to -0.9, indicative of negative price return accompanied by worsening credit spread (thus wrong-way risk, WWR), haircut only increases mildly by 1.13%. Last column of the table shows that haircut changes are negative when the correlation moves to 0.9, indicating right way risk.

---

[12] Note that the default rates are an average of the global corporate default experience between years 1981 to 2014, not necessarily same as S&P's calibrated and official default rates.

[13] Repo on main equity index has gained popularity recently on top of hedge funds' equity financing needs from their prime brokers and/or dealer banks. Bloomberg reports that Soc Gen created a new equity repo strategy to allow investors to gain on equity index repo rate movements, see https://finadium.com/bloomberg-socgen-marketing-quant-equity-repo-index-strategy/.

[14] Currently 'A' rated corporates have an average 5 year CDS at about 125 bp. This level obviously does not apply in a credit contraction cycle.

Table 2. Haircuts for hypothetically 'A', 'BBB', 'BB', and 'B' rated borrowers. DEJD model estimated to 2008-2013 data (Est-1: ($\mu$, $\sigma$, $\lambda$, $p$, $\eta_u$, $\eta_d$) = (0.1231, 0.2399, 79.7697, 0.4596, 169.96, 128.36), Lou 2017), target equivalent wholesale credit rating of 'Aa2', under assumed correlation between equity return and credit spread, and stress market liquidity discounts.

| Borrower CDS | Rating | haircut rho 0 | hc change rho -0.9 | hc change rho -0.9, g 2% | hc change rho 0.9 |
|---|---|---|---|---|---|
| 125 | A | 7.98 | 1.13 | 1.78 | -1.27 |
| 250 | BBB | 9.43 | 1.11 | 1.75 | -1.25 |
| 500 | BB | 10.85 | 1.09 | 1.73 | -1.23 |
| 1000 | B | 12.31 | 1.07 | 1.7 | -1.19 |

The moderate correlation effect is anticipated as the haircut truncated loss exposure lies in the tail of asset returns where the Gaussian component of the asset dynamics is not expected to have a significant impact. Take for example, if the spread volatility is doubled, while keeping the CDS spread at 250, the haircut for a BBB client would increase 0.38% under -0.9 correlation. The general wrong way risk therefore is limited, obviously due to the short MPR and the loss buffer afforded by haircuts.

Specific wrong way risk could occur, if a borrower posts its affiliates' debt instruments as collateral. A structural dependency stronger than the diffusion correlation between asset return and credit spread could be developed. The strongest one in fact is a down jump upon borrower default, which can be modeled as an additional liquidation discount added to $g$. Specific WWR with a 2% asset jump on borrower default has a further increase of haircut of 1.78% (fifth column in Table 2). In this sense, specific WWR is very severe, but the real magnitude will be firm and product dependent, and is best to be analyzed on a case-by-case basis.

Evident from Table 2, haircuts become a tool of credit enhancement to the borrower. A ‘BB’ rated client entering a repo, for example, can post additional 1.42% (=10.85%-9.43%) haircut to make himself a 'BBB' equivalent borrower with the same collateral, both trades showing out an equivalent top credit risk profile of 'Aa2'. For 'BBB', 'BB', and 'B' rated borrowers in the table, the mean reversion speed and spread volatility are kept same but the initial hazard rates are set at 2%, 4.88% and 14.3% to produce 5 year CDS spreads of 250, 500, and 1000 bp respectively[15].

The stylized fact that MMFs are counterparty insensitive when lending to dealers (Krishnamurthy et al 2014, and Hu et al 2019) can be readily explained. Noting that most dealer-banks normally fall into the rating band of 'BBB' and 'A', the separation between 'A' and 'BBB' rated borrowers is small, about 1.45% in haircut. This is not enough a margin for unsophisticated MMFs to invest in their risk management capacities. As a result, in times of stress or crisis, they would opt to other means to restrict lending money to dealer banks, such as shortening repo tenor, reducing lending amount or ultimately shutting down (Comotto, 2012).

In the results presented above, a one-year repo tenor is assumed, as wholesale credit risk management typically standardizes around senior unsecured exposure at one year time horizon. For trades of longer or shorter terms, one can convert it to an equivalent one year credit risk profile to allow comparisons with a firm’s internal credit risk metrics to define haircut targets. Or one could scale the standard one year loss or default rate to the tenor in question based on piecewise constant hazard rates.

Similar collateral haircut tables can be developed for corporates, CMBS, RMBS, and ABS, with proper proxy indices identified and the jump-diffusion model estimated. Results are omitted.

### 3.3. Tri-party and Bilateral Repo Haircut Difference Puzzle

Gorton and Metrick (2012) show dramatic increases of bilateral haircuts during the crisis, while tri-party repo datasets (Krishnamurthy *et al*, 2014, and Copeland *et al*, 2014) show mostly stable haircuts. The abnormally large difference in tri-party and bilateral haircuts, e.g., about 15% for private label MBS prior to Lehman's default and about 30% post default, is not apparent economically. Krishnamurthy *et al* (2014) attribute the rise of haircuts in the bilateral markets to a credit crunch on the part of the dealers, while Copeland *et al* (2014) leave it as an open haircut difference puzzle. Martin *et al* (2014) recognize contributions from different institutional arrangements in that bilateral trades are more custom negotiated, whereas tri-party terms are more template based thus more standardized.

---

[15] These are for illustrative purposes only. An implementation should estimate or calibrate these parameters from single name CDS or properly chosen CDS indices.

Our transaction oriented haircut model offers an explanation of the puzzle from the borrower credit perspective. Dealers as borrowers of cash are generally of much better credit quality than HFs. By looking up from Table 2, if we assume the dealer is 'A' rated, the MMF would charge a haircut of 7.98%. The dealer would charge a 'BB' rated HF 10.85%, assuming zero correlation.

Rating HFs is a very difficult task. Small HFs are highly concentrated in an asset class or a particular kind of investment strategies, so that severe asset valuation distress could threaten the very survival of the fund, an extreme case of significant or specific wrong way risk. Large or multi-strategy HFs organize investment around sub-funds, each of which either pursues a specific strategy or is managed by a portfolio manager (PM). Investment performance is measured at the sub-fund level and investors can choose sub-funds to invest. Sub-funds operate under the parent company's legal umbrella, and as a result they can't be the legal borrowing entity in a repo transaction. But because their investment money and management team are segregated, their use of leverage has to be segregated as well. Repo with a HF is therefore structured to eliminate any credit support from the parent fund. In that sense, it is prudent to apply collateral haircut of 15.53%, resulting in a haircut difference of 7.55%.

Another factor contributing to the difference is the length of the MPR. In the tri-party market the MPR is shorter, because of its institutional efficiency in collateral valuation and settlement (Copeland, Duffie, Martin, and McLaughlin 2012). In the bilateral market, the MPR is generally longer due to trade customization, valuation dispute, and other bilaterally negotiated terms that could prolong the settlement process. Valuation dispute alone may take up to 3 days to resolve. With our model, longer MPR leads to higher haircut. To illustrate, if the MPR drops from 10 days to 5 days, the MMF haircut or tri-party haircut would reduce from 7.98% to 5.34%, widening the haircut difference by 2.64%, to a final number of 10.19%.

Haircut difference between bank and HF counterparties is also evident in Gorton and Metrick (2012)'s bilateral repo dataset, where interdealer bilateral haircuts are compared side by side on the same collateral class with bilateral haircuts facing mid-sized hedge funds ($2-5 billion asset under management), as seen from Table 1 in Dang *et al* (2013). For BBB+/A rated corporate bonds in Jan 2009, for instance, the bilateral haircuts are 0-5% with banks and 35%-40% with HFs,

thus a bank-HF difference of at least 30%. In Jan 2007, the bank-HF difference is much smaller, at about 10%.

### 3.4. Dealer's intermediating liquidity

This bilateral and tri-party haircut difference, if utilized by an intermediating dealer, generates excess cash liquidity. A dealer, for example, can fund its HF client \$84.47 on \$100 security with 15.53% haircut, and gets funded \$94.66 by rehypothecating the security to an MMF at 5.34% haircut, realizing \$10.19 of excess liquidity.

With an analytic model at hand, an upper bound for the excess liquidity can be established. For the first leg where the dealer gets funded by an MMF, the lending MMF would demand a haircut $h_{MMF}$ with the dealer as the counterparty. On the second leg the dealer lends to a HF which can be thought as a counterparty of not much credit worthiness beyond what the collateral asset can afford. In this case, the asset only collateral haircut applies and establishes an up limit of the haircut differential. Table 3 shows the limit for main equities, where $h_{MMF}$ is set to target Moody's 'Aa2' rating with 5 day MPR while the bilateral haircut targets the same rating although at 10 day MPR. Obviously in the setup, the excess liquidity or the haircut differential shown in column 'hc diff' decreases as the CDS spread of the dealer increases, i.e., better credits earn higher excess liquidity.

The excess liquidity could be greater for securitized products, due to higher asset volatility and poorer market liquidity. Comparisons of private label CMO haircuts between triparty and bilateral repo therefore need to be understood with data granularity issues in mind[16].

Note that our computed tri-party haircuts and bilateral haircuts are results of applying the same model to two back-to-back repo trades where the dealer bank is defaultable only when it acts as a borrower. Infante (2019) has a defaultable intermediating dealer in both trades, in order to study an excess liquidity chasing dealer's default impact on the short term funding market. While a broker-dealer default could be a real concern especially during a financial crisis, day-to-day repo transactions have to consider borrower default simply because repo is a credit product.

[16] Usually the triparty repo market finances higher rated private label CMO tranches while the bilateral market have higher percentage of lower rated tranches. Gorton and Metrick (2012) show a rating split of ABS/RMBS/CMBS products into 'AA-AAA' and '<AA', while Krishnamurthy et al (2014) and Copeland et al (2014) have no comparable rating subclass. CMO tranching directly affects haircuts. In Jan 2007 for example, haircuts collected from HFs are 3% for 'AAA' rated ABS papers and 25% for 'BB' rated papers (Table 1 in Dang *et al* 2013).

Table 3. Sample excess liquidity generated in a repo chain with main equity collateral, MPR=10 days. 'hc diff' is the difference between counterparty independent haircut of 15.53% at 10 day MPR and the dealer's tri-party haircuts shown in column '$h_{mmf}$' with 5 day MPR.

| Dealer | Rating | Triparty $h_{mmf}$ | hc diff |
|---|---|---|---|
| 125 | A | 5.34 | 10.19 |
| 250 | BBB | 6.46 | 9.07 |
| 500 | BB | 7.56 | 7.97 |
| 1000 | B | 8.7 | 6.83 |

"Liquidity windfall" (Infante, 2019) gained from the haircut difference in repo rehypothecation is desirable to dealers and is possible prior to the crisis when HFs mostly used only one prime broker. After Lehman's bankruptcy, prime brokerage diversification has taken place. In fact, mega banks' share of prime brokerage businesses have steadily increased and large hedge funds have keen interests in obtaining financing from the banks rather than from dealers who are of much thinner balance sheets. Commercial and investment banks and insurers treat repo as investment products and don't rehypothecate collateral. It is much less convincing that dealers' desire for excess liquidity could be the sole or primary driver behind the haircut differences.

### 3.5. An Empirical Case Study of Repo Haircuts

As historical, transaction level repo haircut data is rare, large scale empirical studies of the model, while desired, are unfeasible at this stage. In this subsection, we conduct a high level empirical case study to see how the model could have predicted haircuts on financing Lehman Brothers during the 2007-2008 financial crisis. Following Gorton and Metrick (2012), we divide the development of the crisis into 4 periods, the first and second halves of 2007 and 2008: 1H2007, 2H2007, 1H2008, and 2H2008. The last trading dates of these four periods are respectively on 6/29/2007, 12/31/2007, 6/30/2008, and 9/12/2008, when Lehman's problem is well publicized and its credit default swaps (CDS) traded last day.

The bank conducts historical data estimation of the DEJD asset model, with a 5 year historical period, including a recent stress period. With new stresses developing during the crisis, the 5 year period is simply the 5 year ending with the last trading day of each period. For instance, 1H2007 period starts on 7/1/2002 and ends on 6/29/2007. Table 4 shows 1, 5 and 10 days 99%

VaR directly estimated from the Bank of America Merrill Lynch's 5 to 10 year average life CMBS 'AA' price return time series. The 10 day VaR doubled from 2H2007 to 1H2008, then nearly quadrupled from 1H2008 to 2H2008.

Having estimated the asset model, the bank considers both the risk neutral log OU model -- fitted to the CDS market on a specific trading day, and the real world model estimated from historical daily CDS curves. Since repo tenors are short, one year in this exercise, there is no need to fit or estimate the full term structure of Lehman's credit curve. For our purposes, we pick the historical 1y CDS spread to regress to estimate the log OU model and bootstrap a default probability curve using only Lehman's 6 month, 1 year and 2 year CDS spreads, which are shown in Table 5 for the last trading days of the four crisis periods. The need for a logarithm model is evident from the multiplying jumps seen in these periods.

Table 4. CMBS ‘AA’ 5 to 10 year average life bond 99% VaR estimated with 5 year historical price return data up to 1st half of 2007, 2nd half of 2007, 1st half of 2008, and 2nd half of 2008, for MPRs of 1, 5 and 10 days. Sharp increases in VaR are observed from 2H2007 to 2H2008.

| VaR(%) | 1-d | 5-d | 10-d |
|---|---|---|---|
| 1H2007 | 0.88 | 1.97 | 2.9 |
| 2H2007 | 1.05 | 2.14 | 3.02 |
| 1H2008 | 1.51 | 3.71 | 6.54 |
| 2H2008 | 3.26 | 9.81 | 23.49 |

Table 5. Lehman Brothers’ short term CDS spreads as of the last trading date of the four cited periods.

| PeriodEnding | 6m | 1y | 2y |
|---|---|---|---|
| 1H2007 | 0.08% | 0.13% | 0.19% |
| 2H2007 | 1.52% | 1.44% | 1.41% |
| 1H2008 | 4.43% | 4.46% | 3.87% |
| 2H2008 | 14.13% | 13.69% | 10.09% |

The estimated log OU model parameters are listed in Table 6. The estimated volatility is quite stable but the mean reversion parameter $k$ becomes negative for 1H2008 and 2H2008, indicating an explosive rather than mean-reverting spread behavior as the broker struggled along the way to final default (yet still a surprise given its 6 month CDS spread is only 14.13%).

Now suppose that the bank adopts a credit policy that targets repo lending at 'Aaa/AAA' rating and applies this policy consistently through the cycle. In 1H2007, asset volatility is low (as reflected from the small VaR in Table 4), borrower credit is good (as indicated by 13 bp of 1 year CDS spread and strong mean reversion), and market liquidity as measured by bid/ask spread is cool. There is not much need for a significant haircut and the asset-only model predicts 5.25% haircut, in line with then BASEL II's supervisory haircut of 8%. With consideration of credit support from Lehman, the risk neutral credit model (assuming zero correlation with the asset return) shows 1.75% haircut, while the historically estimated log OU model results in 2.0%, see last row in Table 7. Another 1.5% could be added to haircut to take into account of the effect of bid/ask related market liquidity.

Table 6. Estimated log OU model parameters using 5 year historical data of Lehman's 1 year CDS spread, for four periods each ending on last trading day of 1H2007, 2H2007, 1H2008, and September 12, 2008, the last day of Lehman's CDS quoted ahead of its September 15 bankruptcy filing.

| | 1H2007 | 2H2007 | 1H2008 | Sep-08 |
|---|---|---|---|---|
| K | 2.4343 | 0.7935 | -0.0212 | -0.3584 |
| σ | 1.2471 | 1.4272 | 1.4618 | 1.4673 |
| $\lambda_0$ | 0.22% | 2.39% | 7.44% | 22.82% |
| λ-mean | 0.13% | 1.44% | 4.47% | 13.69% |

Table 7. CMBS ‘AA’ bond haircuts estimated as the financial crisis unfolds in 2007 and 2008. Column "Fitted" shows haircuts obtained by applying each period's last trading day's CDS curve to the rolling estimated DEDJ model; the "Estimated" column is for haircuts when the historically estimated log OU model is applied to rolling estimated DEJD model; 'Asset-Only' column shows haircuts without consideration of Lehman's credit quality.

| Period Ending | Asset-Only | Fitted | Estimated | Gorton & Metrick | Liquidity |
|---|---|---|---|---|---|
| 2H2008 | 28.5% | 23.2% | 24.2% | 17.1% | 5.0% |
| 1H2008 | 8.7% | 6.1% | 6.7% | 17.1% | 5.0% |
| 2H2007 | 6.4% | 3.7% | 3.7% | 1.8% | 5.0% |
| 1H2007 | 5.3% | 1.8% | 2.0% | 0.0% | 2.0% |

In the second half of 2007, problems in the subprime and mortgage backed securities were well publicized and the bid/ask for ABX.HE's senior tranches increased dramatically, hovering around 5%[17]. At the end of 2H2007, the predicted haircuts are at 3.75% level, although still low by sheer amounts, but roughly doubling the prior period numbers. In 1H2008, predicted haircuts almost double again. And finally when approaching Lehman's final days, model predicted haircuts are close to 25%, in the proximity of the asset-only haircut. Adding in the bid/ask spread, 30% haircut is not inconceivable.

The column labelled "Gorton & Metrick" shows the mean of haircut of Table 2 of Gorton and Metrick (2012), where the 1H2008 and 2H2008 data are shown for the full year 'All of 2008' for 'AA-AAA' ABS/RMBS/CMBS. The average of our model prediction for the full year of 2008 is 14.65% with fitted curve and 15.45% for estimated curve, not too far away from their 17.1% mean haircut.

By applying the haircut model, the bank seems to have been able to respond to the changing market conditions during the crisis. Without availability of comprehensive historical data, this hypothetic case study nonetheless shows the potency of the model.

## 4. Pricing Repo as OTC Derivatives

As discussed earlier, repo is a credit product subject to loss due to default. In the standard reduced form pricing approach, default is modeled as the first jump of a Cox process with a predictable intensity process. Loss at default is determined, and the expected loss is discounted to arrive at the present value (PV) in the risk neutral world[18]. This loss PV is either paid upfront or annualized as a series of premium payments by the counterparty. With the most popular credit

---

[17] There were days when dealers sent out runs showing a bid/ask spread of 10 points. ABX.HE is a series of CDS indices referencing home equity loan backed ABS, essentially those of subprime mortgages.

[18] Consider a T tenor repo, margined at *dt* interval, e.g., one day. Breaking up the repo tenor to a sequence of margin intervals, $t_i$, i=0, 1, 2,..., N. Given a $t_i$, the distribution of loss $L(t_i)$ is computed via equation (11). Let $D(t_i)$ be the applicable discount factor, the net present value (npv) of the loss, or default present value (dpv), is given by,

$$dpv = E[\int_0^T D(t)dL(t)] = D(T)E[L(T)] - \int_0^T E[L(t)]dD(t)$$

where the discount factor is assumed to be independent of the loss. The reverse repo is effectively a floating rate note on a rate index. The index part of the repo interest rate gets back to par when discounted at the same index curve. The present value of a unit spread, or annuity denoted by *apv*, is given by, $apv = \int_0^T D(t)Q(t)dt$, where Q(t) is party C's survival probability and the annuity is computed on a unit notional. The net present value of the repo (npv) is then $npv=1-dpv+S_r*apv$ where $S_r$ is the repo spread. For a repo, dpv is very small due to the presence of significant haircuts and apv is very close to $T$ when $T\leq 1$ year. See Lou (2019) for a review of CDS pricing.

derivative product -- the credit default swap (CDS), for example, the fair swap spread or CDS premium is designed such that its PV offsets that of the loss. In general, the net PV of a credit product is the expected value of the default (or survival) probability weighted cash flows including principal, loss and premium payments. A floating premium leg consists of a reference rate and a spread, commonly perceived to reflect the funding rate and a default risk charge. Repo rates can be quoted either fixed or floating: the former for overnight repos and repos without a reset, and the latter for term repos with at least one reset period.

### 4.1. Repo pricing puzzle

In section 3, we have predicted haircuts targeting Moody's 'Aa2' rating, which carries a one-year expected loss of 0.00075%. This would produce an in-kind repo spread of 0.075 basis point (bp), obviously negligible. Lowering the target rating by a band to 'A2', the expected loss of 0.598 bp remains immaterial. Basically, the repo spread will be close to nil, if all we do is to collect a spread based on expected loss as taught in the standard pricing theory. The median tri-party repo spreads observed are 39, 38, and 20 bps for equity, high yield, and investment grade corporates (Hu *et al* 2019). Repo spreads, therefore, can't be attributed to expected losses. This is what we see as the repo pricing puzzle.

"Haircuts are a puzzle" (Gorton and Metrick 2012), although an old capital market trick invented well before the era of modern finance. Standard finance theories would suggest that "risk simply be priced and the market price reflects risk and risk aversion of the market" (Dan *et al* 2013). Indeed, without the margin period of risk, it is difficult to perceive repo's risk or the need for repo's haircut and repo spread, as the lender can simply liquidate the collateral at the market, instantaneously, without any loss. Consequentially, repo haircuts have to come from other channels, such as uncertainty in collateral value or lack of valuation information (Dan *et al*, 2013).

Recognizing potential repo loss during the MPR has helped us arrive at a haircut model that takes into accounts of factors commonly attributed to haircuts and that seems to be able to produce realistic haircut levels. We also see the institutional aspect of needing haircuts as a credit risk management tool for repo transactions. While we could answer the haircut puzzle by stating that haircuts exist because they are a valuable risk management tool, their comment is still valid in that, as we just demonstrated above, a usual risk driven asset pricing doesn't produce meaningful repo pricing.  Repo is a vanilla credit product, and yet the usual credit analytics fails to price it. This is indeed puzzling.

### 4.2. Economic capital charge in the Black-Scholes-Merton framework

An overnight repo with zero haircut is exposed to gap risk, a risk that the collateral market price jumps overnight. For term repos, traders inherit the term, since the margin period of risk is a matter of days, and the market price moves during the period can't be hedged while the default settlement process is being worked out. This posts an immediate challenge to the celebrated Black-Scholes-Merton options pricing framework and the reduced form risk neutral pricing theory, if we were to price repo as an OTC derivatives or credit products.

Recall that the classic Black-Scholes-Merton economy is complete. An option is attained by dynamic trading in its underlying stock whose discounted price return is a continuous martingale. In other words, the option can be hedged without *any* error. In Merton (1976), stocks' jumps are considered idiosyncratic and not hedgeable. Merton resorts to the hedging portfolio's (infinite) diversification argument under the Arbitrage Pricing Theory (APT) to derive a mixed differential-difference Black-Scholes type equation and an option pricing formula with lognormally distributed jumps. In the credit market, neither perfect replication nor perfect diversification can be done without incurring hedging errors, simply because defaults (as jumps) are known to be strongly correlated (Duffie et al 2007). Therefore there is a systemic, unhedgeable and undiversifiable risk that needs to be bored by a limited number of agents, practically major dealer-banks.

This is a risk clearly missing in these traditional finance theories used in the industry. It is from this systemic nature of hedging errors that Lou (2016, 2019) reckons that economic reserves would be required and charging customers for its use as a compensation necessary. The practice of including a capital charge on the side for complex derivatives and structured products that reflects a firm's shadow cost of capital has been around for long, even prior to the financial crisis. With the arrival of the post-crisis regulatory risk capital framework reforms, some researchers in the industry have worked to formalize such a charge into derivatives pricing. A quick and practical response has been to incorporate the strengthened regulatory capital requirements into existing derivatives pricing and valuation models, to facilitate dealer banks' transfer cost of capital. The regulatory capital based KVA (capital valuation adjustment, Green, Kenyon and Dennis, 2014) has been controversial, however, as others argue that cost of capital is a cost of doing business rather than a measure of risk and valuation. Prampolini and Morini (2018) frame KVA as a by-product of regulatory constraints[19] that force conservative market hedges.

Lou (2016, 2019) takes a different approach in considering KVA based on *economic* capital which links to systemic hedging errors. To briefly recap, in Lou (2016), a repo is placed into a self-financed derivative economy where it is hedged with the underlying security. Let π denote the wealth of the economy, and $\Gamma_t$ the default indicator function of the borrower, the hedging error is shown to be

$$-dA_t = d\pi_t - r\pi dt = -d\Gamma_{t-u} l(\tau + u) + (1 - \Gamma_t) El(t)\lambda dt \qquad (15)$$

where τ is the default time, *u* the default settlement lag, i.e., MPR, *r* the (nominal) risk free rate, λ the default intensity of the borrower. *l(t)* is the loss on the repo, and *El(t)* is the expected loss.

The potential default loss during the MPR (eqt. 1) is treated as a gap risk, unheadgeable and to be warehoused. A reserve capital account is set up in the economy that requires a rate of return at the shadow cost of capital. The delta hedging strategy is designed to zero out mean hedging error while a capital reserve is taken as its VaR measure. Specifically, we have

$$\hat{\pi}_t = \pi_t - \frac{\beta_T}{\beta_t}\pi_T$$

$$VaR_t = \inf\{x \in R : \Pr(\hat{\pi}_t > x) \le 1 - q\}. \qquad (16)$$

where $\beta_t = e^{-\int_0^t r ds}$ is the usual deflator on the risk free rate *r*. The risk neutral pricing formula for the fair value of the repo *v(t)* is then shown to be

$$\upsilon(t) = E_t\left[\int_t^T e^{-\int_t^s r_e du}\left(s_p N_p(s) - s_k N_c(s) - \lambda El(s)\right)ds\right] \qquad (17)$$

where *T* is the repo tenor, $r_c$ is the senior unsecured financing rate of the borrower, $N_p$ is the repo principal amount, assumed to be 1 here, $N_c$ is the reserve (economic) capital amount, $s_p$ the repo spread defined as the difference between repo rate $r_p$ and the cost of fund $r_f$, $s_p = r_p - r_f$, and $s_k = r_k - r$ is the capital charge spread with $r_k$ being the shadow cost of capital.

---

[19] Additional capital buffers, RWA (risk weighted asset), and the Leverage Ratio are factors often cited that have impacted post-crisis business practice.

For an at-the-issue repo, its fair value is zero. This leads to a repo break even formula,

$$r_p = r_f + s_k N_c + \lambda El \tag{18}$$

Adding a profit margin to the break-even rate then arrives at street traders' repo rate.

In application, $El$ and $N_c$ (=EC) are provided in section 2, since the effect of the deflating in equation (16) is negligible as repo tenors are short and the residual hedging error is essentially same as (1.b).

Because haircuts affect economic capital, this pricing model establishes an implicit link between haircuts and repo rates. In fact, plug in equation (14.b), we see explicitly that the repo rate is negatively linear in haircut.

$$r_p = r_f + s_k C_0 - s_k L_{gd} h + \lambda El \tag{18.b}$$

### 4.3. Repo transaction pricing example

Table 8 shows a three month repo trade with a hedge fund counterparty and a portfolio of liquid US stocks. We use SPX500 as the proxy for the collateral. With trade haircut of 8%, we use estimated SPX500 to compute repo's EL and ES. The risk charge as annualized EL is only 1 bp, but ES is 2.39% which translates into a capital charge of 48 bp when applied with a shadow cost of capital at 20%. The final repo spread consists an internal cost of fund of 35 bp and desk's margin of 40 bp.

Table 8. Sample trade pricing worksheet: risk charge (EL) and capital charge for a fund borrower, assuming correlation to SPX500 at -0.9, 10 day MPR, q=99.9%.

| Trade | | Pricing | |
|---|---|---|---|
| Principal (mm) | 25 | Borrower 5y CDS (bp) | 750 |
| Tenor (m) | 3 | Asked HC (%) | 8 |
| Pay/reset | Monthly | Risk Charge (bp) | 1 |
| Collateral Type | US Main Equity | Capital Charge (bp) | 48 |
| Counterparty | Hedge fund | Desk Mark-up | 40 |
| Floating Index | LIBOR 1 month | Cost of Fund | 35 |
| Evergreen | 3m/2m/3m | Final | 124 |

As a part of the pricing exercise, we can plot the risk and capital charges as a function of haircuts in Figure 2. The capital charge far dominates the risk charge. The linear decline in the capital charge is evident, as governed by equation (14.b and 18.b).

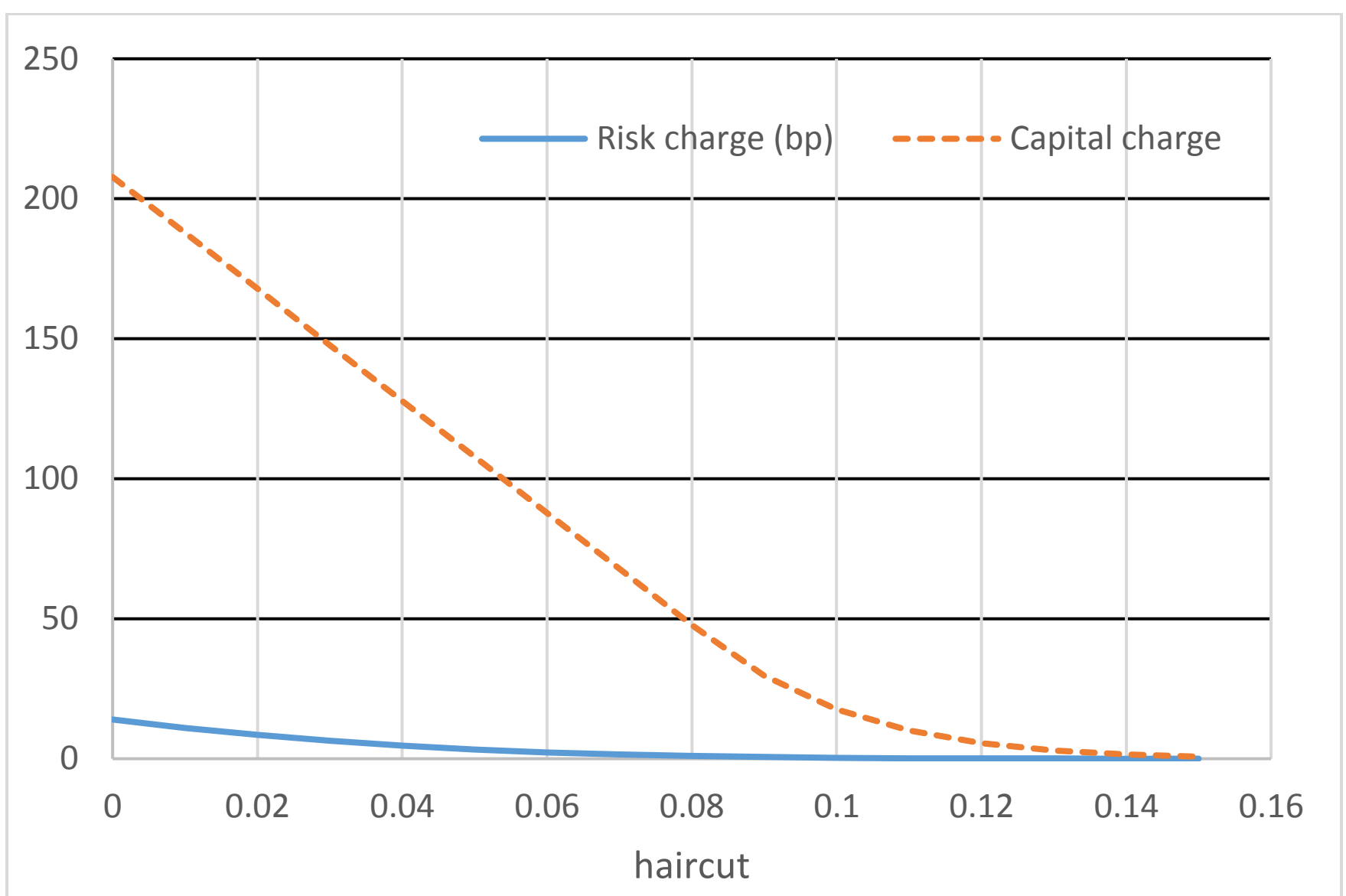

Figure 2. Sample risk charge (= EL/T) compared to capital charge (=ES*20% RoE) as haircuts change, with MPR 10 days.

### 4.4. Optimal repo pricing

While a dealer bank can weigh in its haircuts and repo rates offerings, this dual pricing measure offers clients a chance to negotiate as they always do. For an end-user such as a hedge fund, its primary economic motivation in a repo transaction is the funding rate, the all-in rate that covers both the fund provided for from the repo and the residual funding need due to the application of a haircut. Since an end-user typically does not have access to the unsecured borrowing market, the residual fund has to come from its equity or investor capital. Therefore the fund's return target or shadow cost of capital is a proper rate to apply. This leads to a simple all-in rate formula,

$$r_a = (1-h) * r_p + h r_c \tag{19}$$

where $r_c$ is the ROE (target return on equity capital), and $r_a$ is the all-in rate.

If the repo rate is a constant, the all-in rate would be linear in haircut and has a minimum at $h=0$, since $r_c > r_p$, indicating the end-client would have preferred to transact at zero haircut, to minimize his overall funding cost. Needlessly to say, this is not happening (exactly because haircuts are subject to separate approval and risk management), with the exception of overnight Treasuries repos where the theoretical haircut is so small and its impact can be ignored.

Plugging equation (18.b) into (19) leads to a dominantly quadratic term in repo haircut $h$. The all-in rate therefore has a minimal point, as plotted in Figure 3.The fair repo rate steadily declines as haircut increases but is floored at the cost of fund rate. The all-in rate declines first when haircut increases up to about 7.5%, reaches a minimal of 2.3692%, then increases as haircut goes up further, reflecting the end-user's cost on funding the residual part increases at a faster pace.

The repo pricing formula thus offers a mechanism of repo trade negotiation through variation of haircuts. End-users can lock a minimal overall funding rate based on the dealer-bank's pricing. Haircut and repo rate as two pricing measures of repo transaction work as a tandem, with the client's all-in rate optimization providing the pivot of the tandem.

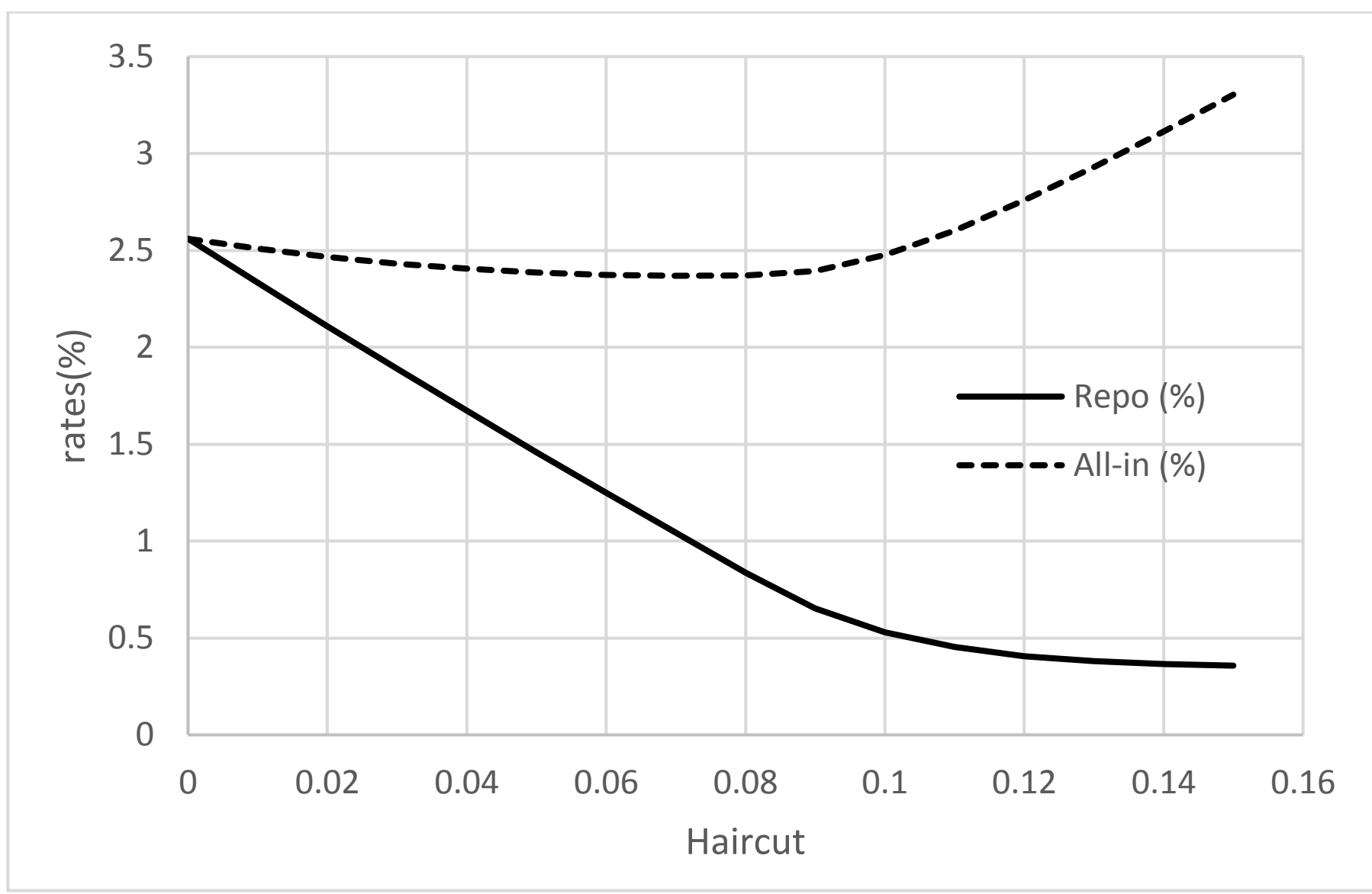

Figure 3. Break-even repo rate and a client's all-in rate vary as haircut changes. The all-in rate has a trough or minimal of 2.3692% at haircut of 7.5%.

## 5. Economic Capital Driven Haircuts

In section 3, we have shown haircuts designed to meet certain high rating based on traditional PD (aka S&P method) or EL (aka Moody's method) criteria. Given economic capital's prominent role in repo pricing shown in section 4, it is natural to ask whether haircuts should be designed based on EC criteria. This is relevant in that there are circumstances when securities are used as collateral without any compensation, i.e., there is no way to charge a fee or interest for a risk profile same as a repo. Collateral for bilateral OTC derivatives, initial margin and variation margin for central counterparty clearing facilities are ready examples. Such is the case, a logical decision is to design haircuts such that the resultant risk profile during the margin period of risk bears negligible pricing significance.

### 5.1. Haircuts targeting EC

Taking out cost of fund from equation (18), the repo spread becomes sum of the risk charge and the capital charge. Since the capital charge dominates the risk charge, an agent may opt to control the haircut to minimize the capital charge or equivalently EC. This is accomplished by equation (5). In Figure 2, the level of haircut needed to control ES charge at 1 bp is 14.6%.

As discussed earlier, counterparty independent collateral haircuts are used solely to mitigate counterparty credit risk, so that a pricing equation does not need to exist. EL and EC calculations can proceed, however, without the credit support from the counterparty (Lou 2017). So superficially, one can still apply equation (18) for the purpose of haircut setting.

Table 9 lists superficial risk and capital charges on US main equities with haircuts at 0, 5, 10, and 15%. The first row shows that 12% haircut would alone produces 1 bp risk charge, but not total charge. The second row has capital charge based on VaR, and the last row is based on ES. If the standard is set at CVaR 99.9% as is in BASEL 3, the haircut needs to be 14.8% or17.5% with CVaR and ES used as EC respectively.

Noting that CVaR 99.9% drops to zero at 14.8% haircut, we can use zero as the hurdle in equation (5) when CVaR is used as the EC measure.

Table 9. Minimum haircuts (in column "min HC(%)") needed to control charges below 1 bp and comparison of risk charge (row "EL-only") with combined risk charge and capital charge ("CVaR + EL" where CVaR is used for EC and "ES+EL" where ES is taken as EC.)

| | min HC(%) | Chrg (bp) @15% hc | Chrg (bp) @10% hc | Chrg (bp) @5% hc | Chrg (bp) @0% hc |
|---|---|---|---|---|---|
| EL-only | 12 | 0.11 | 3.48 | 39.49 | 196.7 |
| CVaR + EL | 14.8 | 0.11 | 55.96 | 177.98 | 542.4 |
| ES+EL | 17.5 | 12.11 | 68.96 | 190.98 | 554.4 |

### 5.2. EC as a market stabilizing tool

Repo haircut is known to be cyclical (CGFS 2010, Comotto 2012). When credit is in an up-cycle, asset volatility is low, borrower credit is improving, and market liquidity is abundant and smooth, lower levels of haircuts seem appropriate. When times are bad, things are reversed with haircuts raised and liquidity squeezed, and deleverage could lead to downward spiral, resulting in an unstable funding market.

This procyclical effect has led the revised regulatory framework (CGFS 2010, BCBS 2020) to stipulate a counterparty insensitive and stable haircuts through credit cycles. Whether a stable through the cycle haircut is the best policy and/or business tool to ensure repo funding market stability is debatable. Traders would argue that repo's tenors are shorter than the time it historically takes to develop a market-wide stress so that fixing haircut levels according to distressed market experience is counter-economic and punitive to repo trades. Surveyed bilateral repo haircuts do go below and above BASEL's supervisory haircuts, as seen from Martin *et al* (2014) and Gorton and Metrick (2012).

Now that EC has a dual role in haircut setting and repo spread pricing, it can be used as an effective tool for funding market stabilization purposes. In a credit expansion cycle, a desk can agree to a (lower) haircut at the current market condition, and charges (higher) economic capital cost into the fair value of the repo trade, thus deterring the leveraging-up effect. In a down cycle, haircuts move higher, but the repo pricing would be lower as the economic capital and thus its capital charge reduces, along with anticipated lowering of cost of fund, the net result is slowing down deleverage and preventing repo runs.

The economic capital we referred to is different from regulatory capital (RC). In the collateral haircut approach to counterparty exposure, BASEL III first determines an equivalent wholesale exposure as $(E\text{-}M(1\text{-}hvol))^+$ where $E$ is exposure or principal in a repo, $M$ is collateral market value, and *hvol* is BASEL's volatility adjustment *aka* haircut. For a repo traded at haircut $h$ such that $E=M(1\text{-}h)$, the exposure becomes $M(hvol\text{-}h)^+$. For regulatory purposes, it needs to estimate loss given default (LGD) and PD (probability of default) of an unsecured exposure of its counterparty. The borrower's PD and LGD and repo's tenor (floored at 1 year) are input to BASEL's wholesale credit risk capital requirement formulae to compute regulatory capital.

We compute and plot both EC and RC in Figure 4 for a one-year repo trade with a hypothetic 'BBB' rated counterparty on US main equities collateral. Based on BASEL III, the repo has no exposure when the haircut is greater or equal to the supervisory market price volatility of 15%, assuming 10 day MPR. When haircut is less than 15%, the difference becomes the exposure which is applied with a risk weight of 191% assuming 3.08% PD and 60% LGD for the counterparty with a stressed correlation factor of 1.25 and a multiplier 1.06 applied. The regulatory capital increases linearly from zero to 2.3% at zero haircut. The economic capital calculated at 99.9 percentile ES has a maximum of 6.58% at zero haircut, about three times of the regulatory capital. Note that ES drops to zero around 15% haircut where the probability of no loss is greater than 99.9%.

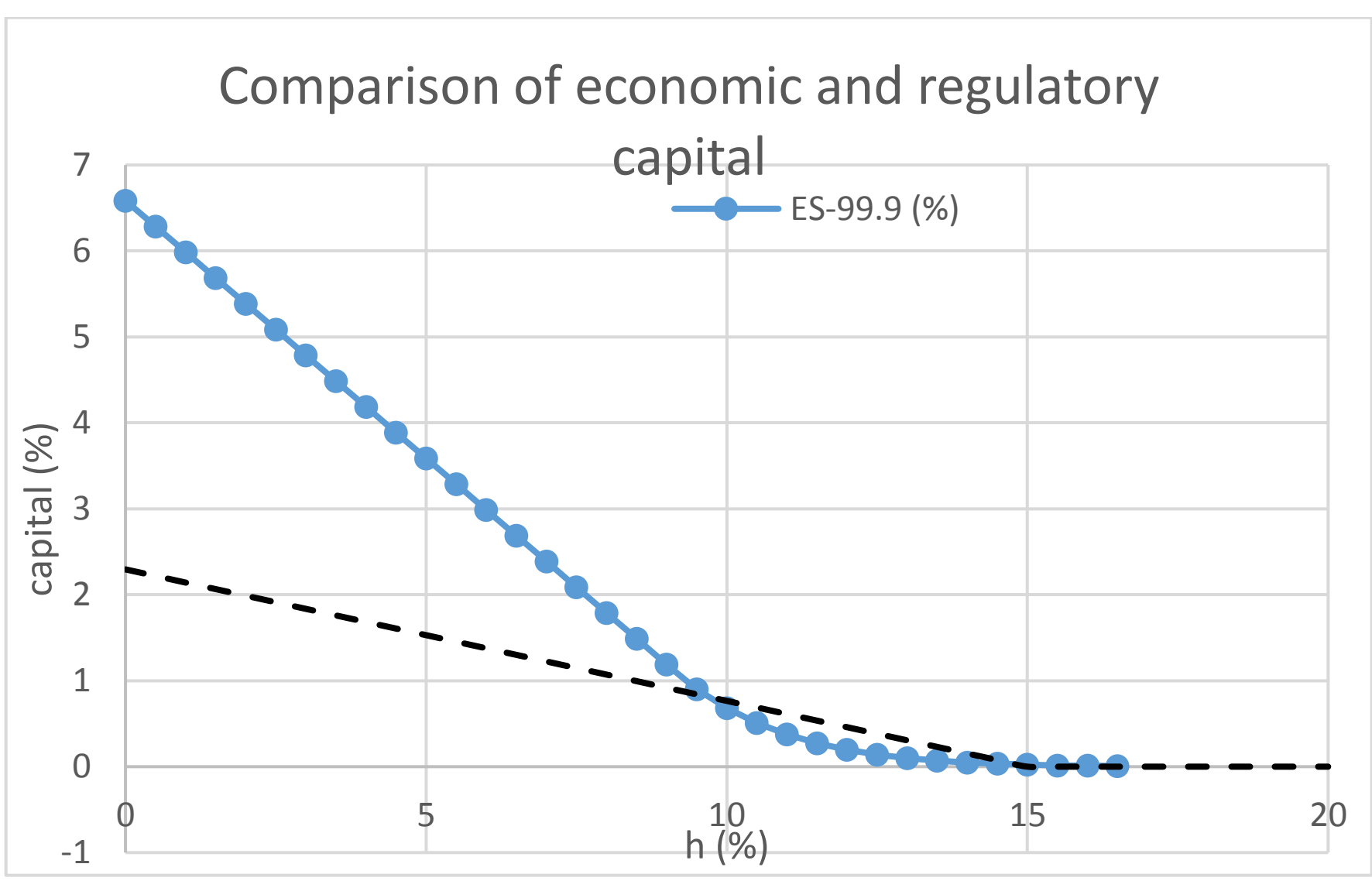

Figure 4. Comparison of economic capital and regulatory capital for a sample 1 year repo trade with a 'BBB' counterparty (with 5 year CDS at 250 bp) with SPX main equity collateral under zero correlation, 10 day MPR.

Figure 4 shows that regulatory capital overestimates economic capital in the range of 10% to 15% haircut and underestimates it when trade haircut moves further away (left and smaller) from the 15% cutoff. Comparing to the real economic capital, the regulatory capital basically penalizes low risk trades (when haircuts are high) and awards high risk (when haircuts are low.) Relying solely on regulatory haircuts to charge cost of capital is dangerous.

**5.3. Economic capital and maturity compression**

Maturity compression (Krishnamurthy *et al* 2014) or "flight from maturity" (Gorton, Metrick and Xie 2014) is a phenomenon where MMF lenders tend to shorten the tenors of their lending during a time of great uncertainty. One way to deal with foreseeable uncertainty is to raise the haircuts but to keep the same tenor, but apparently they have chosen the alternative of reducing the tenor while keeping the same haircut. Exactly what risk or risk measure are they averse to?

Suppose that dealer banks A and B adopt EL and EC based haircut definitions respectively. Bank A targets Moody's 'Aa1' rating and charges 9.04% haircut to a top quality dealer bank on a one year term repo with US main equities collateral. Bank B targets on ES at q=99.9% and will have to charge a much higher 15.92% haircut, assuming 10% ROE used to convert 'Aa1' target EL rate to EC target. If the repo tenor is reduced to 1 month, Bank A basically sees the same (EL based) haircut, while bank B sees a reduced 12.24% haircut.

Had bank B calibrated its ES target such that it also sees the same 9.04% haircut for the one year repo, applying the new ES target to 1 month repo would result in a haircut of only 2.62%. Figure 5 plots this maturity effect on haircuts for the tenor range of 10 days to 250 days. As shown, EL based haircuts ("HC-EL-10" for 10 day MPR and "HC-EL-5" for 5 day MPR) are flat in the first half of the tenor range and slightly inclined due to the CDS's upward sloping term structure from 6 month to 1 year. EC based haircuts ("HC-ES-10" for 10 day MPR and "HC-ES-5" for 5 day MPR), however, show a clear tendency of compacting on shorter tenors and steeper decreases with shorter tenors.

Now suppose a market stress is developing. The greater uncertainty would imply a higher haircut of 9% at 250 day maturity. If the client is only willing to give 5% haircut, to stay at the same level of risk tolerance as measured by EC, the bank would slide down the curve, say, “HC-ES-10”, to the lesser maturity of about 50 days. By sliding down on the tenor while keeping the same haircut therefore reflects a strategy of keeping an option of early exit as a response to heightened uncertainty.

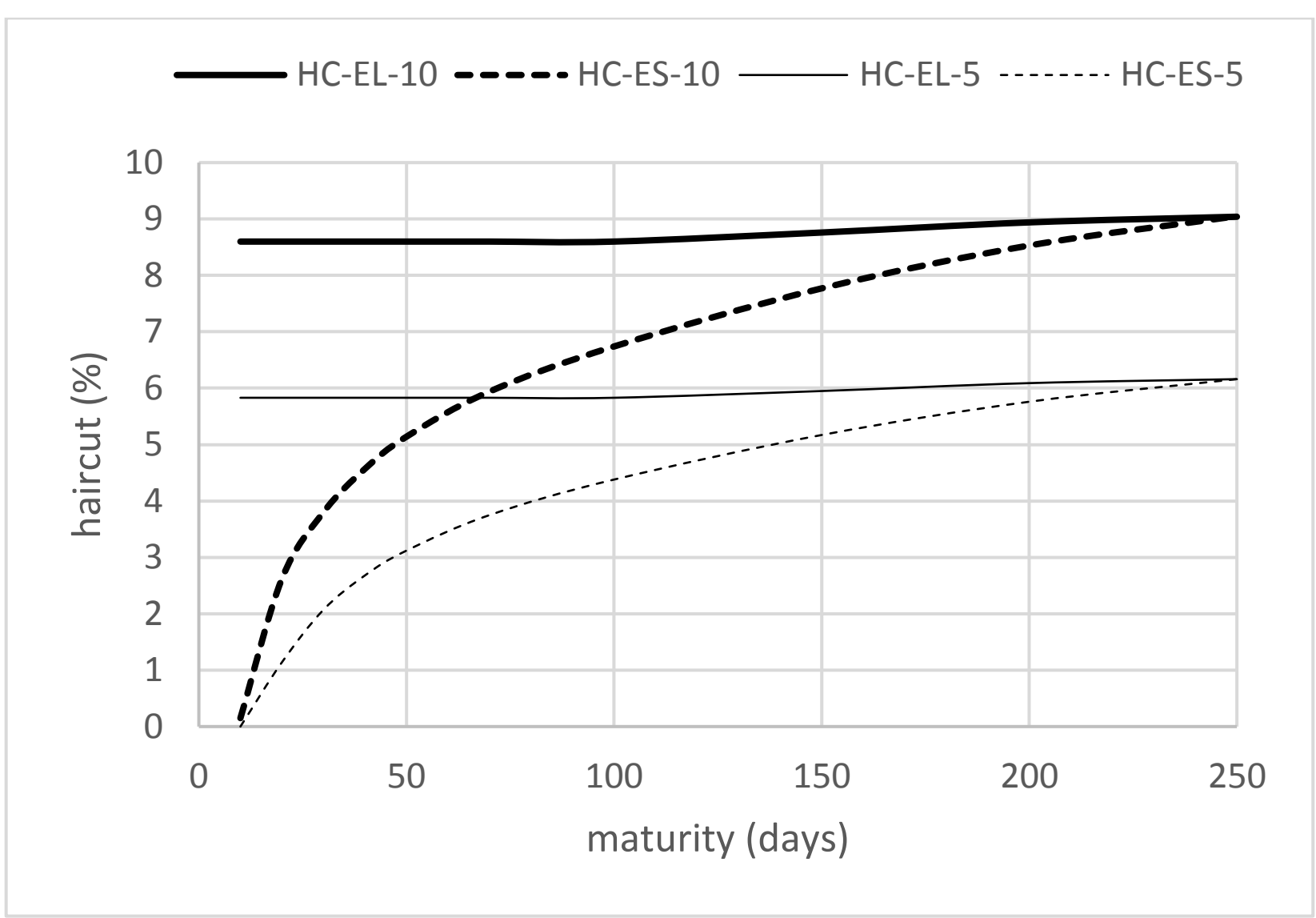

Figure 5. Haircuts vary with repo tenors when EL target and EC/ES target are used, for a counterparty of a spot CDS curve (6m, 1y, 2y) = (47.5, 61.3, 75.3) bps. 'HC-EL-10' is haircuts via EL definition with 10 day MPR targeting 'Aa1' rating, 'HC-ES-10' via EC/ES with 10 day MPR normalized such that its ES target produces the same haircut for 1y term repo as EL target. 'HC-EL-5' and 'HC-ES-5' are the same but with 5 day MPR.

Indeed, maturity difference between tri-party and bilateral segments is prevalent. Krishnamurthy *et al* (2014) and Copeland *et al* both state that 90% of tri-party repos their dataset cover are overnight or open repos, with maximum term of 7 days. In the bilateral market, the median term reported by Auh *et al* (2016) is 10 days for government securities and 30 days otherwise, as loan maturity tends to increase as collateral quality drops.

## 6. Conclusion

As a simple and old debt product, repos possess a puzzling dual pricing measure: the co-existence of haircuts and repo spread. We treat haircuts as a credit risk management tool that seeks to control lender's exposure to collateral price gap risk. Haircuts are determined so that the credit risk profile of a repo achieves certain performance criteria, such as maximum expected loss or probability of default given rating targets (e.g. S&P's 'AA+' or Moody's 'Aa2') or minimal economic capital (e.g. one year, 99.9-percentile credit risk VaR).

Our transaction repo haircut model incorporates asset volatility and jump risk, borrower credit risk, correlation, and market liquidity risk. As is expected of normal circumstances, predicted haircuts are primarily driven by collateral asset risk and market liquidity. The model can reproduce empirically observed elevations of bilateral haircuts in asset backed securities following the subprime crisis. Haircuts are only weakly dependent on counterparty credit and its asset correlation. Because active dealers' credit quality are in a close proximity, applicable haircuts when they borrow are thus largely insensitive to dealers' identity, a fact well documented in the tri-party repo market.

When dealers intermediate between hedge funds and money market funds, their superior credit quality than hedge funds' results in a lower haircut charge to dealers and a higher haircut for hedge funds. Taking the hedge fund's credit quality to the extreme as if there is no-recourse, the haircut difference then sets a limit on excess liquidity a dealer could gain by intermediating. The tri-party market's superior settlement mechanism, in particular, its elimination of collateral pricing disputes, affords a shorter margin period of risk than the bilateral market. These explain the haircut difference between tri-party and bilateral repo markets.

For actively traded repos, haircuts are negotiated. When traded haircuts are lower, a trader naturally expects to price up repo rates to compensate. If he applies industry standard risk neutral credit pricing models, he will be disappointed as repo expected loss is very small and produces negligible repo spreads. We find that the key to the so-called repo pricing puzzle is to recognize the jump or gap nature of repo risk. Since gap risk is neither hedgeable nor diversifiable, a capital reserve is required and a capital usage charge follows. Repo spread in fact is dominated by the compensating capital charge. Take for example, if a one year repo on US main equities is traded with a 'BBB' rated borrower at 7.5% haircut, the expected loss is 0.2 bp while the economical capital charge is 24 bp.

A break-even repo rate formula is derived by pricing repo as an OTC derivatives in a Black-Scholes-Merton framework, where the hedging error's value-at-risk or tail loss is taken to measure economic capital. It consists of a cost of fund component, a risk charge representing the expected repo loss, and a capital charge at the shadow cost of capital. In particular, economic capital is negatively linear in repo haircuts, so is the fair repo spread. Repo haircuts and repo spread as a dual therefore work like a tandem.

From the borrower's perspective, the all-in funding rate allows an optimal choice of haircuts that balances, and settles uniquely between repo's dual pricing measures. The tandem allows a repo to be priced in a naturally counter-cyclical manner, a solution to address repo haircuts' procyclicality and repo funding market instability induced thereafter. In a credit expansion cycle, for example, traded repo haircuts tend to be low, but lower haircuts lead to higher economic capital and higher capital charges in repo pricing, thus deterring an excessive build-up of leverage.

When a market stress is developing with heightened uncertainty, repo haircuts are expected to rise. For borrowers wanting to maintain the same leverage (aka the same haircut), the model requires a reduced repo tenor, to control the economic risk via EC, explaining the observed tendency of lenders shortening repo tenors. The model could aid future econometric studies of repo haircuts, repo pricing and their determinants, which will be strongly desired once relevant transaction data is available.

## Appendix A. Proof of Haircut Loss

The loss function defined in equation (1) is driven by a kernel function $H(t,u) = \left(\frac{1-h}{1-g} - \frac{B_{t+u}}{B_t}\right)^+$, where $g$ is a constant liquidity premium and $h$ is a constant haircut. Let $X(t,u) = \frac{B_{t+u}}{B_t}$. *X(t,u)-1* is one period simple return of the security.

For an arbitrary $h^*$ satisfying $h^*>h$, rewrite the indicator function as,

$$I\left[X(t,u) \leq \frac{1-h}{1-g}\right] = I\left[X(t,u) \leq \frac{1-h^*}{1-g}\right] + I\left[\frac{1-h^*}{1-g} < X(t,u) \leq \frac{1-h}{1-g}\right] \quad \text{(A.1)}$$

And the loss kernel as follows,

$$H(t,u) = \left(\frac{1-h}{1-g} - X(t,u)\right) I\left[X(t,u) \leq \frac{1-h}{1-g}\right] = \left(\frac{1-h^*}{1-g} - X(t,u)\right)^+ + \frac{h^*-h}{1-g} I\left[X(t,u) \leq \frac{1-h^*}{1-g}\right] + \left(\frac{1-h}{1-g} - X(t,u)\right) I\left[\frac{1-h^*}{1-g} < X(t,u) \leq \frac{1-h}{1-g}\right] \quad \text{(A.2)}$$

Let $b = \frac{h^*-h}{1-g}$, $b>0$ since $h^*>h$, and $H^*(t,u) = \left(\frac{1-h^*}{1-g} - X(t,u)\right)^+$, then we have,

$$H(t,u) = H^*(t,u) + bI\left[X(t,u) \leq \frac{1-h^*}{1-g}\right] + \left(\frac{1-h}{1-g} - X(t,u)\right) I\left[\frac{1-h^*}{1-g} < X(t,u) \leq \frac{1-h}{1-g}\right] \quad \text{(A.3)}$$

Now consider random variable $I[H(t,u) > b]$ for a given pair of *(h,b)*. Define $h^* = h + b(1-g)$. Obviously,

$$I[H(t,u) > b] = I\left[X(t,u) \leq \frac{1-h^*}{1-g}\right] = I[H^*(t,u) > 0] \quad \text{(A.4)}$$

Therefore, the kernel cumulative distribution satisfies,

$$\Pr(H(t,u) > b) = E[I[H(t,u) > b]\,] = \Pr(H^*(t,u) > 0) \quad \text{(A.5)}$$

This could be made more general as follows,

$$\Pr(H(t,u) > b) = \Pr(H^*(t,u) > b^*) \quad \text{(A.6)}$$

where $h^* + b^*(1-g) = h + b(1-g)$, $b \geq 0$ and $b^* \geq 0$.

Because the loss function $L$ is a simple function of the kernel, if *F(x|h)* denotes the cumulative loss distribution with haircut $h$, then *F(b*|h*)= F(b|h)*.

Next consider the expectation of the truncated loss, *E[L(t,u)| L(t,u)>l]*. Noting (A.4),

$$H(t,u) * I[H(t,u) > b] = H(t,u) * I[H^*(t,u) > 0] = H^*(t,u) + bI[H^*(t,u) > 0]$$

$$E[H(t,u)|H(t,u) > b] = \frac{E\left[H(t,u)*I[H(t,u)>b]\right]}{\Pr(H(t,u)>b)} = \frac{E[H^*(t,u)]}{\Pr(H^*(t,u)>0)} + b \qquad \text{(A.7)}$$

Applying the constant to get the expected truncated loss,

$$E[L(t,u)|L(t,u) > l] = \frac{E[L^*(t,u)]}{\Pr(L^*(t,u)>0)} + l, \qquad \text{(A.8)}$$

$$l = (1 - R_c)(1 - g)M_0 b$$

In the above, we have not considered counterparty default during the repo tenor of *T*. Counterparty default as a single trigger event could happen at a small time interval prior to *T*, but the magnitude of loss is still determined by the margin period immediately following the default. The resultant loss distribution will need to integrate over the length of *T* but the integration does not change the equality established in A.6 and A.8.